\documentclass{PoS}

\newcommand{\be}{\begin{equation}}
\newcommand{\ee}{\end{equation}}
\newcommand{\beq}{\begin{eqnarray}}
\newcommand{\eeq}{\end{eqnarray}}
\newcommand{\bea}{\begin{eqnarray*}}
\newcommand{\eea}{\end{eqnarray*}}

\usepackage{epsfig}
\usepackage{epsf}
 \usepackage{dsfont}
\usepackage{slashed}
\usepackage{booktabs}
\usepackage{graphicx}
\usepackage{float}

\newcommand{\Op}{\mathcal{O}} 
\newcommand{\PO}{\mathcal{P}} 
\newcommand{\J}{\mathcal{J}}
\newcommand{\Dlr}{\buildrel \leftrightarrow \over D\raise-1pt\hbox{}}
\title{Hadron Structure and Form Factors}

\ShortTitle{Hadron Structure and Form Factors}

\author{{C. Alexandrou}\\ 
  Department of Physics,  University of Cyprus, P.O. Box 20537, 1678 Nicosia,
Cyprus and Computation-based Science and Technology Research Center,
Cyprus Institute, P.O. Box 27456, 1645 Nicosia, Cyprus\\
        E-mail: \email{alexand@ucy.ac.cy}}

\abstract{ We review recent results on hadron form factors and 
nucleon generalized parton distibutions
obtained with dynamical lattice QCD simulations. 
We discuss  lattice artifacts  and open questions, and present the connection
of lattice results to hadron structure and to the corresponding
quantities measured in experiment.
}

\FullConference{The XXVIII International Symposium on Lattice Field Theory, Lattice2010\\
		June 14-19, 2010\\
		Villasimius, Italy}

\begin{document}

\section{Introduction}
Lattice QCD simulations are currently being performed  with  dynamical  degenerate u- and d- quarks with a mass close  to their physical value 
as well as  the strange quark, using a number of different
discretization schemes with the most common being 
Wilson-improved, staggered and chiral fermions.
Furthermore, simulations at several lattice spacings and volumes are 
becoming available, enabling a comprehensive study of lattice artifacts.
The masses of low-lying hadrons have been computed and extrapolated to
 the continuum limit using large enough lattice sizes to   ensure
 that volume effects are small~\cite{Durr, Alexandrou}. These calculations show agreement with experiment
and therefore pave the way for evaluating other phenomenologically interesting  quantities beyond these masses.

\begin{figure}[h]
\begin{minipage}{0.53\linewidth}
\hspace{-0.2cm}{\includegraphics[width=\linewidth]{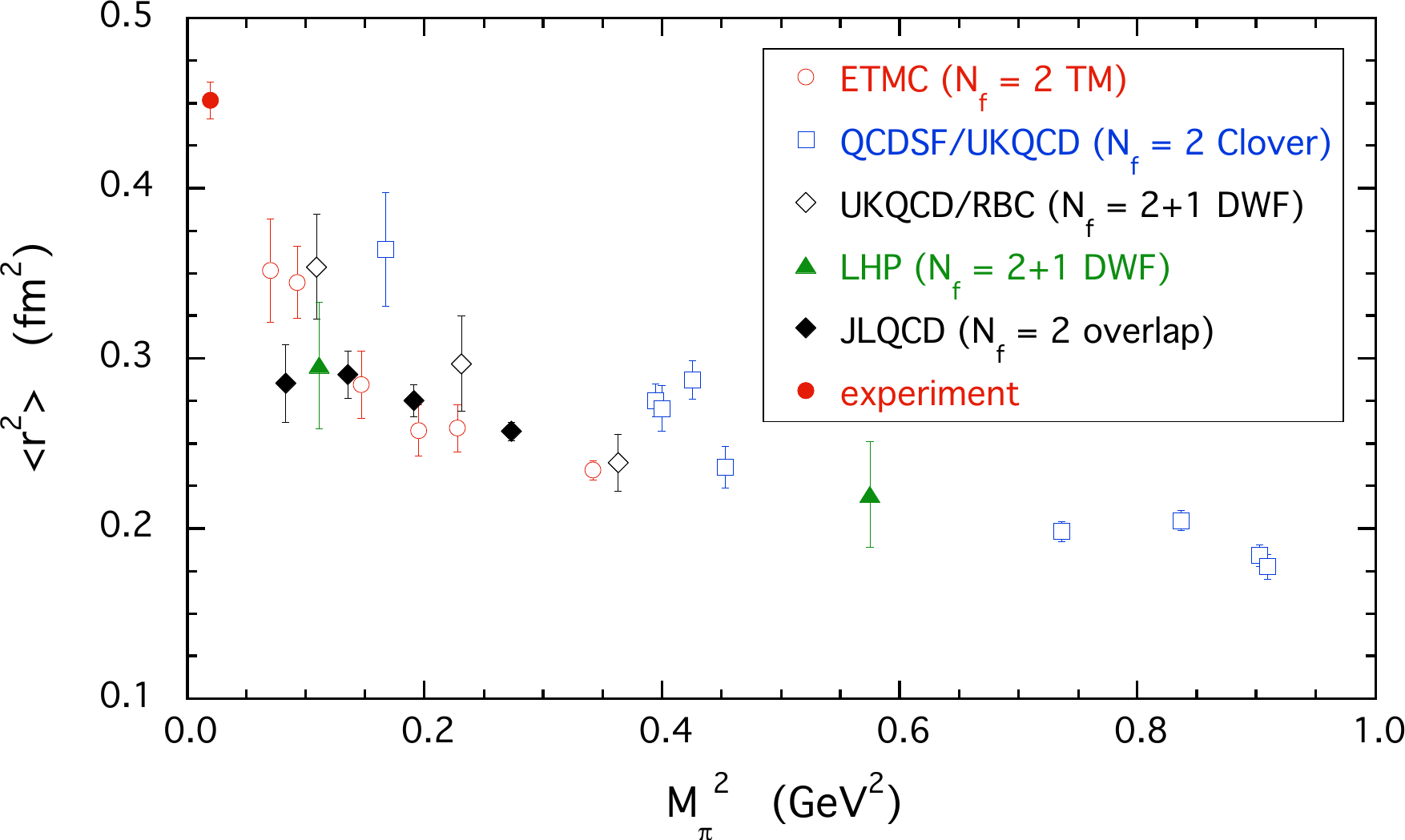}}
\caption{The pion mean square radius as a function of $m_\pi^2$ obtained using simulations with $N_f=2$ twisted mass quarks.}
\label{fig:r2_pion}
\end{minipage}\hfill
\begin{minipage}{0.43\linewidth}
     {\includegraphics[width=0.9\linewidth, height=0.8\linewidth]{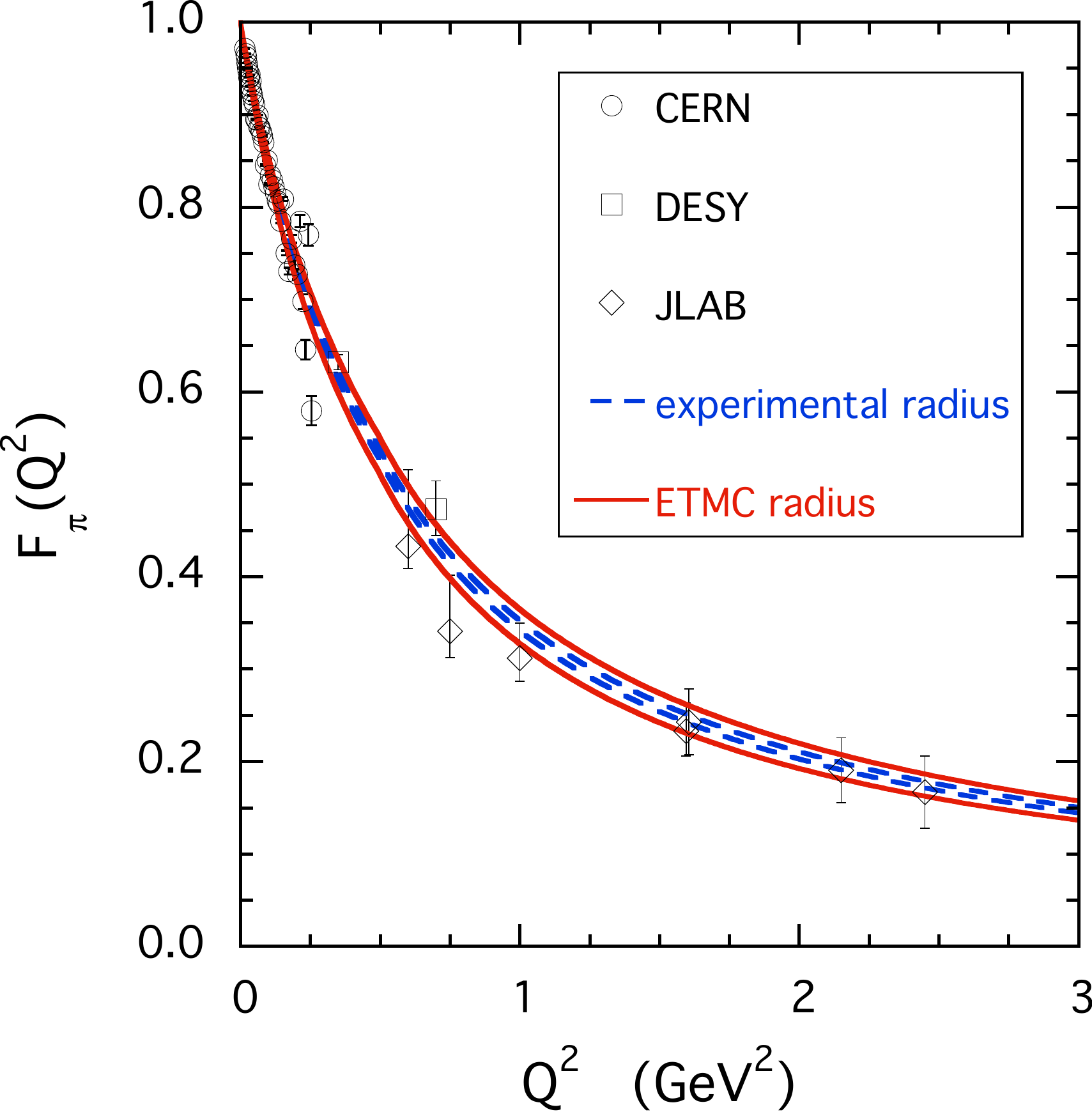}}
\vspace*{-0.3cm}\caption{$F_\pi$ extrapolated to the physical pion
mass (red band) using NNLO chiral PT compared to experiment (blue band).}
\label{fig:fpi}
\end{minipage}
\end{figure}

Several collaborations, using dynamical quarks with pion mass down to about
 300~MeV, have calculated the pion electromagnetic (EM) form factor~\cite{fpi}, which is obtained from the matrix element
 $\langle \pi^+(p^\prime) |J_\mu| \pi^+(p) \rangle =(p_\mu+p^\prime_\mu) F_\pi(q^2)$, where $q^2=(p^\prime-p)^2=-Q^2$. Based on
vector dominance, lattice data are fitted to the form  $F_\pi(Q^2)=\left(1+\langle r^2 \rangle Q^2/6\right)^{-1}$ to extract the mean squared radius, which
is shown in Fig.~\ref{fig:r2_pion}. As can be seen, there is an 
increase in the value of   $\langle r^2\rangle$ at small pion mass, $m_\pi$. An accurate
extraction of $\langle r^2\rangle$ benefits from evaluating the form factor
at small values of $Q^2$ accomplished by using twisted boundary conditions (b.c.).
In a recent calculation, ETMC combined twisted b.c.
and the so  called `one-end' trick to incorporate
the all-to-all propagator and
improve statistics. Using simulations with two degenerate light quarks ($N_f=2$)  at two lattice spacings and two volumes~\cite{ETMC}
  the assessment of cut-off and volume effects was carried out.
 Lattice results on $F_\pi$ obtained with
 pion masses  in the range of 300~MeV to 500~MeV, are extrapolated to the physical point using NNLO chiral
perturbation theory (PT). The resulting form factor is shown in Fig.~\ref{fig:fpi}~\cite{ETMC} and it is in agreement with experiment.

\begin{figure}[h]
\begin{minipage}{0.45\linewidth}\vspace*{-0.5cm}
      {\includegraphics[width=1.1\linewidth]{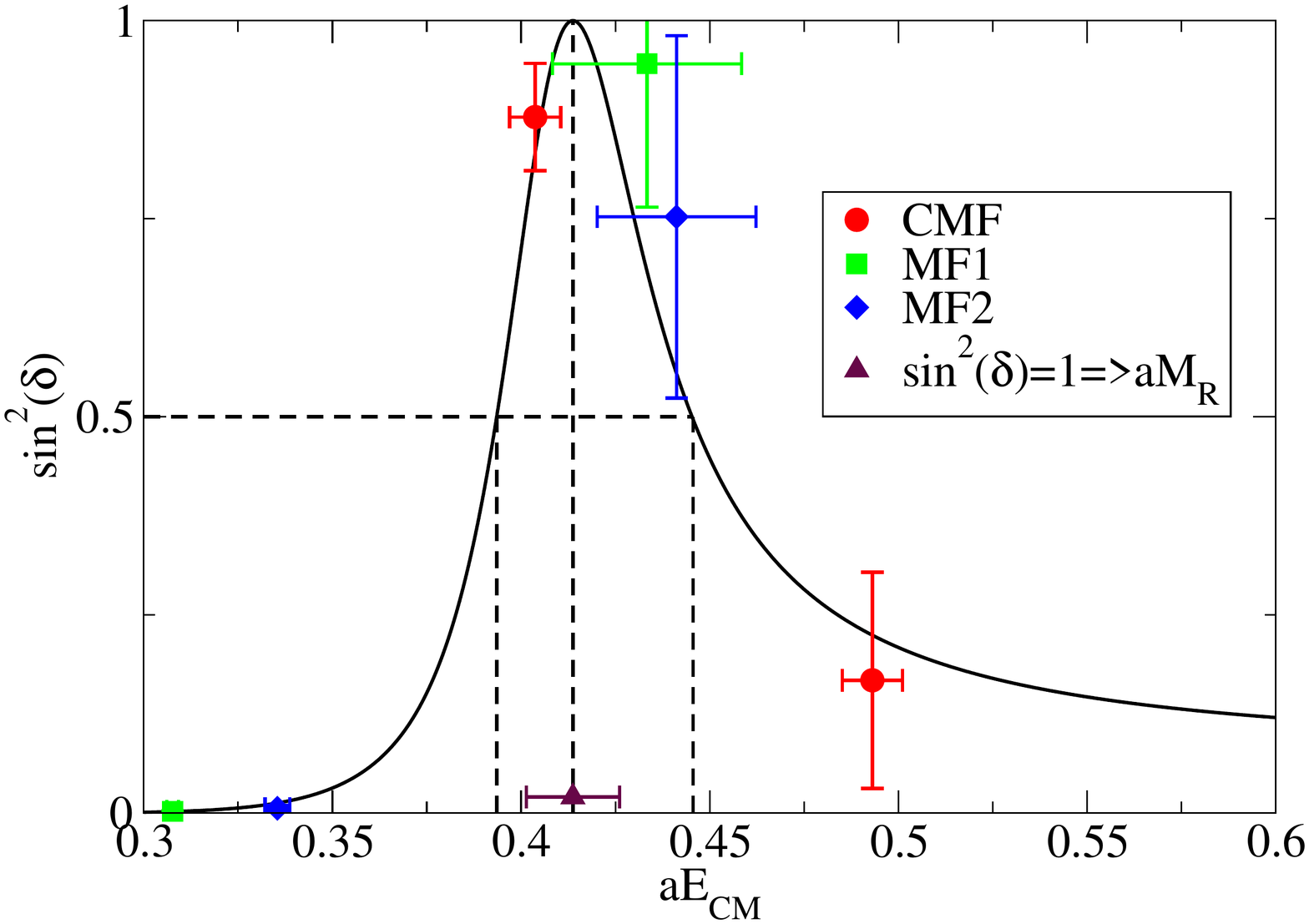}} \vspace*{-1cm}
\caption{The $\rho$-meson phase shift at $m_\pi=308$~MeV for a lattice of  $L=2.8$~fm.}
\label{fig:delta rho}
\end{minipage}\hfill
\begin{minipage}{0.45\linewidth}\vspace*{-0.5cm}
      {\includegraphics[width=1.1\linewidth]{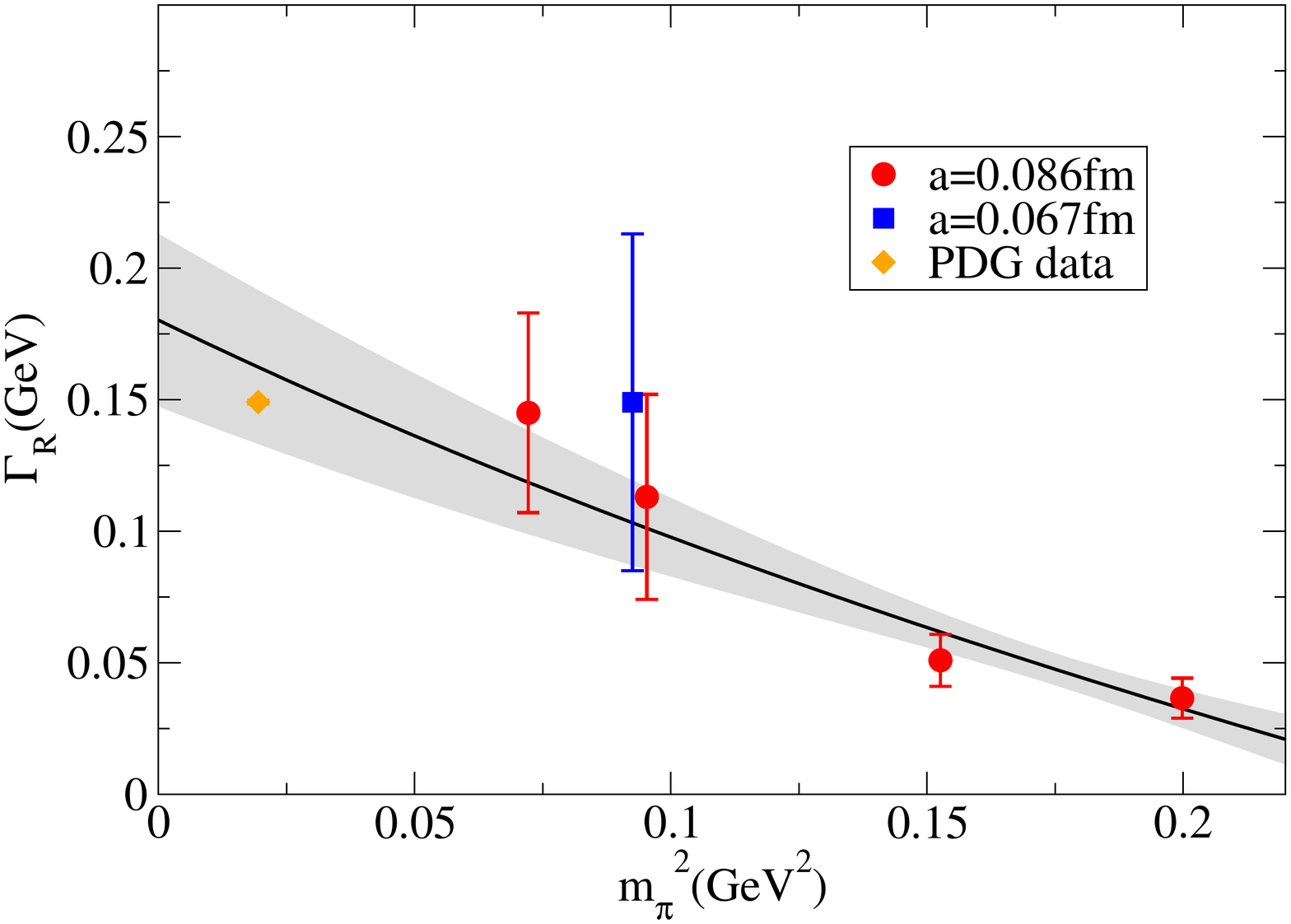}}\vspace*{-0.5cm}
\caption{The $\rho$-meson width for $N_f=2$ twisted mass fermions as a function of $m_\pi^2$.}
\label{fig:rho width} 
\end{minipage}
\end{figure}

As simulations with quark masses close to the physical value become available, the study of resonances and decays of unstable particles becomes an important issue. 
The $\rho$-meson width has  been studied by several groups~\cite{rho-width}. Considering a
 $\pi^+\pi^- $ system  in the $I=1$-channel, the P-wave scattering phase shift $\delta_{11}(k)$ in infinite volume is related 
via L\"uscher's relation  to the energy shift in a finite box.
Using $N_f=2$ twisted mass fermions (TMF) and considering the  center of mass frame and two moving frames one extracts the phase
shift at different values of the energy, shown in Fig.~\ref{fig:delta rho}. 
From the effective range formula $\tan \delta_{11}(k)= \frac{g^2_{\rho\pi\pi}}{6\pi}\frac{k^3}{E_{CM}\left(M^2_R-E_{CM}^2\right)}$,
where $k=\sqrt{E_{CM}^2/4-m^2_\pi}$   one  determines $M_R$ and the coupling $g_{\rho\pi\pi}$ and then extracts the width using $\Gamma_\rho=\frac{g_{\rho\pi\pi}^2}{6\pi}\frac{k_R^3}{M^2_R}$, where $k_R=\sqrt{M_R^2/4-m^2_\pi} $.
The results on the width as a function of $m_\pi^2$ are shown in Fig.~\ref{fig:rho width}~\cite{Feng}.

Having reproduced the  low-lying hadron spectrum~\cite{Durr, Alexandrou, LHPC, PACS},  the masses of  
excited states can be studied  using e.g. variational 
methods~\cite{excited}. Furthermore, one can go beyond masses and
consider form factors (FFs) and
generalized parton distributions (GPDs) that probe hadron structure.
The characterization of nucleon structure, in particular, is considered a milestone in hadronic 
physics and many experiments have been carried out to measure nucleon 
FFs and structure functions. Experiments on nucleon FFs started in the 50s. 
A new generation of experiments using polarized beams and targets
 are yielding high precision data spanning larger $Q^2$ ranges. Therefore,
nucleon FFs  serve as a further benchmark for lattice QCD.
FFs provide ideal probes of  the charge and magnetization distributions of the 
hadron as well as a determination of its shape in analogy to similar studies in e.g. deuteron and other nuclei.

\begin{minipage}{0.54\linewidth}\vspace*{-3cm}
\hspace*{-0.7cm}Non-relativistically the form factor can be related\\
\hspace*{-0.7cm} to the density distribution via \\
$F(\vec{q}^2)=\int d^3x e^{-i\vec{q}.\vec{x}}<\psi|\rho(\vec{x})|\psi>$. \\
\hspace*{-0.7cm}In Fig.~5 we show the intrinsic charge density contours \\
\hspace*{-0.7cm}of a spin-zero nucleus showing deformation revealed \\
\hspace*{-0.7cm}through measurements of transition densities using\\
\hspace*{-0.7cm}electron scattering.
\end{minipage}
\begin{minipage}{0.45\linewidth}\vspace*{-0.3cm}
\hspace*{1cm}{\includegraphics[width=0.65\linewidth]{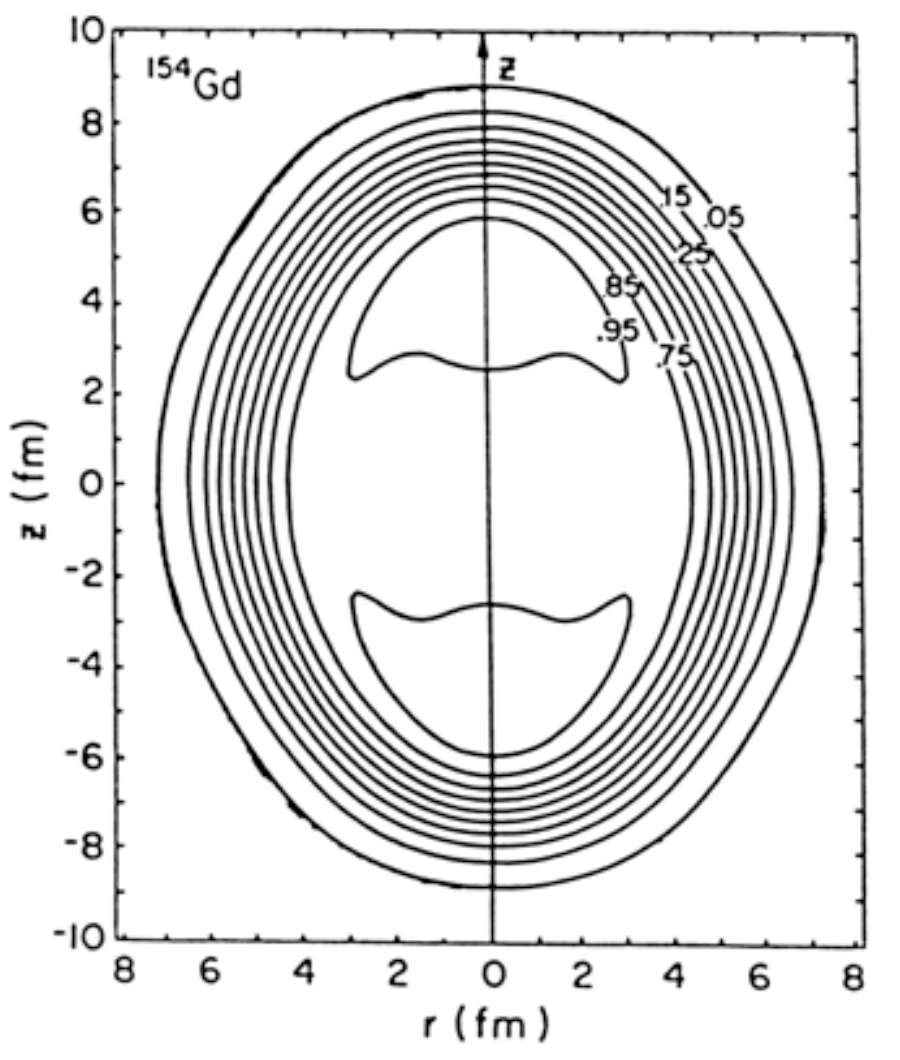}} \\
{\small {\bf Figure 5:} Tomographic view of the zero-spin  deformed nucleus $^{154}Gd$ derived from its rotational bands  using electron scattering.}
\vspace*{0.3cm}
\end{minipage}\hfill

In what follows we will review the status of lattice QCD calculations on baryon form
factors and nucleon generalized parton distributions.

\section{Nucleon Generalized Form Factors - Definitions}
In this section, we briefly define the quantities for which results are presented. 
High energy scattering can be formulated in terms of light-cone correlation functions.
Considering one-particle states $|p^\prime \rangle $ and $|p\rangle$,  GPDs are defined by~\cite{Diehl, Ji}:
\vspace*{-0.5cm}

$$
  F_{\Gamma}(x,\xi,q^2) =   \frac{1}{2}\int \! \frac{d\lambda}{2\pi}
     e^{ix\lambda} \langle p^\prime |\bar {\psi}(-\lambda n/2) {\Gamma}
\PO      e^{ig\!\int \limits_{-\lambda /2}^{\lambda /2}\! d\alpha n \cdot A(n\alpha)}
     \psi(\lambda n/2) |p\rangle \,,
 $$
where 
$\overline{P}=(p^\prime+p)/2$, $\xi=-n\cdot q/2$, $x$ is the momentum fraction, and $n$ is a light-cone vector with $\overline{P}\cdot n=1$. 
\vspace*{0.3cm}

\begin{minipage}{0.7\linewidth}
\hspace*{-0.8cm}There are three different types of operators, depending on the
choice of $\Gamma$.\\
\hspace*{-0.8cm}Considering nucleon states these are   
\begin{eqnarray*}
 \Gamma &=&\slashed{n}: \rightarrow \frac{1}{2}\bar{u}_N(p^\prime)\left[\slashed{n}{ H(x,\xi,q^2)}+i\frac{n_\mu q_\nu \sigma^{\mu\nu}}{2m_N}{ E(x,\xi,q^2)}\right] u_N(p) \\
\Gamma & =&\slashed{n}\gamma_5 : \rightarrow   \frac{1}{2}\bar{u}_N(p^\prime)\left[\slashed{n}\gamma_5 {{\tilde{H}(x,\xi,q^2)}}+\frac{n. q \gamma_5}{2m_N} {\tilde{E}(x,\xi,q^2)}\right] u_N(p)\\
\Gamma&=& n_\mu\sigma^{\mu\nu}  : \rightarrow {\rm tensor \,\, GPDs} 
\end{eqnarray*}
   \end{minipage}\hfill
\begin{minipage}{0.3\linewidth}\vspace*{-1.cm}
\vspace*{1cm}\hspace{0.5cm} ``Handbag'' diagram \\ \vspace*{-0.3cm}
\hspace{-0.3cm} \includegraphics[width=\linewidth]{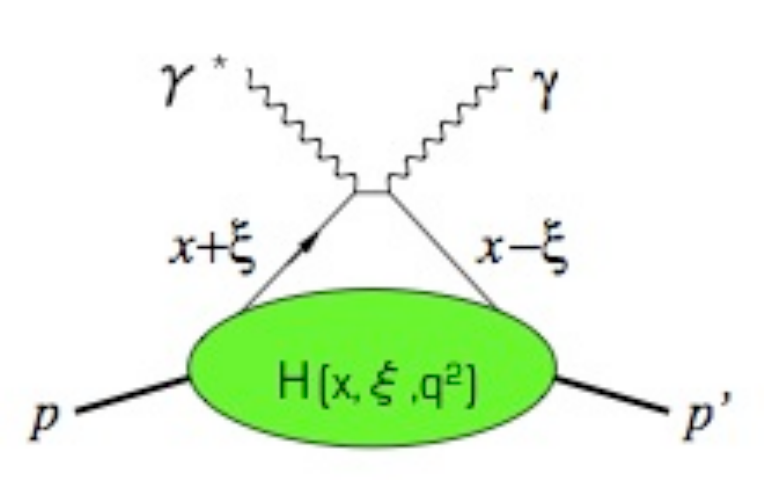}
 \end{minipage}

\vspace*{0.3cm}
\noindent
 Expansion of the light cone  operator leads to a tower of local twist-2 operators $\Op_\Gamma^{\mu\mu_1\ldots\mu_{n}}$, related to moments. The
diagonal proton matrix elements 
$\langle P|{\cal O}_\Gamma(x)| P\rangle$, measured in
deep inelastic scattering, are  connected to the   parton distributions $q(x)$, $\Delta q(x)$,  $\delta q(x)$. The twist-2 operators are defined by

\vspace*{-0.3cm}

   \begin{eqnarray*}
   \Op_\slashed{n}^{\mu\mu_1\ldots\mu_{n}}    = 
  \bar \psi  \gamma^{\{\mu}i\Dlr^{\mu_1}\ldots i\Dlr^{\mu_{n}\}} \psi : &\stackrel{unpolarized}{\rightarrow}&
\langle x^{n}\rangle_q = \int_{0}^{1}dx \, x^n\left[q(x)-(-1)^{n}\bar{q}(x)\right] \> \\
  \tilde\Op_{\slashed{n}\gamma_5}^{\mu\mu_1\ldots\mu_{n}}      = 
\bar \psi  \gamma_5\gamma^{\{\mu}i\Dlr^{\mu_1}\ldots i\Dlr^{\mu_{n}\}} \psi : & \stackrel{helicity}{\rightarrow}&
\langle x^n\rangle_{\Delta q} = \int_{0}^{1}dx \, x^n\left[\Delta q(x)+(-1)^{n}\Delta\bar{q}(x)\right]  \\
   \Op^{\rho\mu \mu_1\ldots\mu_{n}}_{n_\mu\sigma^{\mu\nu}} = 
\bar \psi  \sigma^{\rho\{\mu}i\Dlr^{\mu_1}\ldots i\Dlr^{\mu_{n}\}} \psi  : &\stackrel{transversity}{\rightarrow} &\langle x^n\rangle_{\delta q}=\int_{0}^{1}dx\, x^n\left[\delta q(x)-(-1)^{n}\delta\bar{q}(x)\right] 
   \end{eqnarray*}\vspace*{-0.3cm}

\noindent
where
$ q=q_\downarrow+q_\uparrow, \Delta q=q_\downarrow-q_\uparrow, \delta q=q_\top+q_\perp $, and the curly brackets represent a symmetrization over indices and subtraction of traces.
The off-diagonal matrix elements extracted
from deep virtual Compton scattering can be written in terms of generalized form factors (GFFs), which contain both form factors and parton distributions:
 \beq {\langle N(p^\prime,s^\prime) |} \Op_\slashed{n}^{\mu\mu_1\ldots\mu_n} {| N(p,s) \rangle} &=& 
    \bar{u}_N(p^\prime,s^\prime) 
     \Biggl[ \sum_{i=0,{\rm even}}^{n}\left( {A_{n+1,i}(q^2)} \gamma^{\{\mu}+{B_{n+1,i}(q^2)}\frac{i\sigma^{\{\mu \alpha}q_\alpha}{2m_N} \right) q^{\mu_1}\ldots q^{\mu_{i}} \Biggr.\nonumber\\
 &{}& \hspace*{-0.3cm} \overline P^{\mu_{i+1}}\ldots\overline P^{\mu_{n}\}}+\Biggl.{\rm mod}(n,2) {C_{n+1,0}(q^2)} \frac{1}{m_N} q^{\{ \mu}q^{\mu_1}\ldots q^{\mu_{n}\}} \Biggr] u_N(p,s)
\eeq
  and similarly for $\Op_{\slashed{n}\gamma_5}$ (in terms of $\tilde A_{ni}(q^2)$, $\tilde B_{ni}(q^2)$) and $\Op_ {n_\mu\sigma^{\mu\nu}}$ (in terms of 
  $A_{ni}^T,\ B_{ni}^T,\ C_{ni}^T$ and $D_{ni}^T$).
 We list the following special cases:\\
$\bullet$  $n=1$: Ordinary nucleon form factors:
\bea 
           &{}&A_{10}(q^2) = F_1(q^2)=\int_{-1}^1 dx\, H(x,\xi,q^2), \quad B_{10}(q^2) = F_2(q^2)=\int_{-1}^1 dx\, E(x,\xi,q^2) \nonumber \\
           &{}&  \tilde A_{10}(q^2) = G_A(q^2)=\int_{-1}^1 dx \,\tilde{H}(x,\xi,q^2), \quad \tilde B_{10}(q^2) = G_p(q^2)=\int_{-1}^1 dx \,\tilde{E}(x,\xi,q^2)\, ,
\nonumber \eea 
where in the case of the EM current,  $j_\mu=\bar{\psi}(x)\gamma_\mu \psi(x)$, the nucleon matrix element is written in the form
      $\bar{u}_N(p^\prime,s^\prime)\left[\gamma_\mu { F_1(q^2)} + \frac{i \sigma_{\mu\nu} q^\nu}{2 m_N} F_2(q^2)\right]u_N(p,s)$.
The Dirac {$F_1$} and Pauli {$F_2$}  FFs are  related to the electric and magnetic Sachs FFs via the relations:
  ${G_E(q^2)} = {F_1(q^2)} - \frac{q^2}{(2m_N)^2}{F_2(q^2)} $ and
     ${G_M(q^2)} = {F_1(q^2)} + {F_2(q^2)}$.
For the axial vector current $A^a_\mu=\bar{\psi}(x)\gamma_\mu\gamma_5 \frac{\tau^a}{2}\psi(x)$  the nucleon matrix element is of the form
$ \bar{u}_N(p^\prime,s^\prime)\left[ \gamma_\mu\gamma_5 { G_A(q^2)} + \frac{q_\mu\gamma_5}{2 m_N} {G_p(q^2)} \right] \frac{1}{2}u_N(p,s)$.\\
$\bullet$  $A_{n0}(0)$, $\tilde A_{n0}(0)$, $A^T_{n0}(0)$ are moments of 
parton distributions, e.g. $\langle x \rangle_q   = A_{20}(0) $ and 
$\langle  x \rangle_{\Delta q} = \tilde A_{20}(0)$ are the spin 
independent and helicity distributions. Knowing these quantities one can  
 evaluate the quark spin,  $J_q = \frac{1}{2}[ A_{20}(0) + B_{20}(0)]=\frac{1}{2}\Delta \Sigma_q+ L_q$ and investigate the fraction of the spin carried by quarks and its contribution to the total spin via the 
 nucleon spin sum rule,
$\frac{1}{2} = \frac{1}{2}\Delta \Sigma_q+ L_q+J_g$, as well as 
the momentum fraction
carried by gluons via
           the  momentum sum rule: $\langle x \rangle_g = 1-A_{20}(0)$.

\section{Lattice evaluation}
In order to extract the matrix elements connected to GFFs
we need to evaluate three-point correlators and compute the renormalization of
the operators involved.
Despite recent progress on the evaluation of 
disconnected loops, most lattice
calculations of GFFs do not take  into account disconnected contributions.
Therefore, in what follows, we consider iso-vector operators for which   such  contributions are
zero in the isospin limit. 
For one-derivative operators, mixing with lower dimension operators is avoided by
 symmetrizing over the Lorentz
indices and making them traceless.
The study of  cut-off and finite volume effects in a systematic
way has just begun for baryon GFFs.
These are more difficult to assess   since
chiral expansions that describe such dependencies are   not as developed as in the light meson case. The presence of more uncertainties in
the chiral expansion combined with the  larger statistical noise, which
for the nucleon two-point function increases like $\frac{\rm noise}{\rm signal}\sim  e^{(m_N-3m_\pi/2)}/\sqrt{N}$,
 make the extrapolation of these quantities
 to the physical point much more demanding.
In this review we will focus on:
i) Nucleon form factors and lower moments using dynamical simulations with pion mass $m_\pi\stackrel{<}{\sim} 500$~MeV and spatial lattice length
 $L\stackrel{>}{\sim} 2$~fm 
and ii) the $N$-$\Delta$ system in order to determine the complete set of coupling constants needed in chiral expansions.
 Other topics relevant to hadron structure, such as the
 strange nucleon FFs, hyperon, Roper and nucleon negative parity FFs, distribution amplitudes and transverse momentum dependent parton distributions can be found in Ref.~\cite{Zanotti} and in 
contributions to this volume.

$\bullet$ {\bf Three-point functions:} 
For the extraction of matrix elements of local operators we need the evaluation of two-point and three-point functions defined by

\begin{minipage}{0.33\linewidth}
   \begin{eqnarray}     
    &{}& \hspace*{-1.15cm}G(\vec p, t) =\sum_{\vec x_f} \, e^{-i\vec x_f \cdot \vec p}\, 
     {\Gamma^4_{\beta\alpha}}\, \langle J_{\alpha}(\vec x_f,t_f){\overline{J}_{\beta}(0)} \rangle \nonumber\\
    &{}&\hspace*{-1.15cm} G^{\mu\nu}({\Gamma},\vec{p}^\prime,\vec q, t) =\sum_{\vec x_f, \vec x} \, e^{i\vec x \cdot \vec q}\, e^{-i\vec x_f \cdot \vec p^\prime}\, 
     {\Gamma_{\beta\alpha}}\, \langle {J_{\alpha}(\vec x_f,t_f)} {\cal O}^{\mu\nu}(\vec x,t) {\overline{J}_{\beta}(\vec x_i,t_i)} \rangle .\nonumber
   \end{eqnarray}
\end{minipage}\hfill
\begin{minipage}{0.32\linewidth}\vspace*{-0.8cm}
  \hspace*{-0.3cm}   \includegraphics[width=1.3\linewidth]{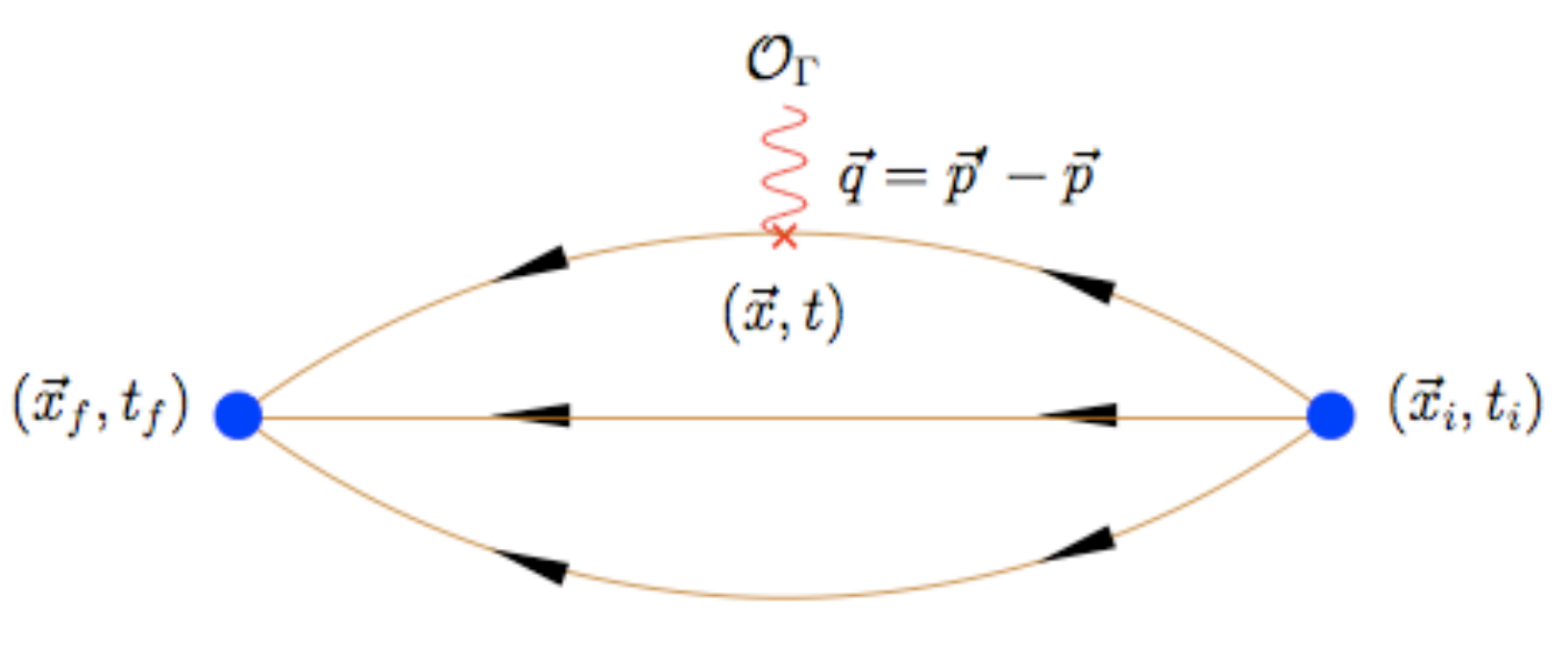}
   \end{minipage}

\noindent
Only  the displayed connected diagram is evaluated, which, for
most current applications, is done 
by using sequential inversion ``through the sink'' fixing the sink-source separation  $t_f-t_i$, final momentum $\vec p^\prime $ and $\Gamma$- projection matrices.
Smearing techniques  are crucial for improving ground state dominance in three-point  correlators and thus keep $t_f-t_i$ as short as possible. We stress that
it is important to ensure that the time separation $t_f-t_i$ used is sufficiently large by performing
the calculation at a bigger time separation and checking the consistency of the results. 
The generalized eigenvalue method can
further improve  identification of the ground state~\cite{Blossier}
and enlarge the upper range of accessible $Q^2$-values~\cite{Lin}.

$\bullet$ {\bf Renormalization constants:}
Most collaborations use non-perturbative renormalization. Using  a momentum dependent source~\cite{Gockeler} one evaluates
\be S^u(p) = \frac{a^8}{V}\sum_{x,y} e^{-ip(x-y)} \left\langle u(x) \bar u(y) \right\rangle \,,\,\,\, G(p) = \frac{a^{12}}{V}\sum_{x,y,z,z'} e^{-ip(x-y)} \langle u(x) {\bar u(z) \J(z,z') d(z')} \bar d(y) \rangle
\ee
 with the amputated vertex functions given by
$\Gamma(p) = (S^u(p))^{-1} \ G(p)\ (S^d(p))^{-1}$ and $\J$ determines the operator,
e.g. $\J(z,z') = \delta_{z,z'} \gamma^{\{\mu}\Dlr^{\nu\}}$ would
correspond to the local vector current.
The Z-factors can be  determined  in the RI$^\prime$-MOM
scheme by imposing the following conditions
\be
   Z_q = \frac{1}{12} {\rm Tr} [(S(p))^{-1}\, S^{(0)}(p)] \Bigr|_{p^2=\mu^2}  \,,\,\,\,\,\,\,
   Z_q^{-1}\,Z_{\cal O}\,\frac{1}{12} {\rm Tr} [\Gamma_{\mu\nu}(p) \,
\Gamma^{(0)-1}_{\mu\nu}(p)] \Bigr|_{p^2=\mu^2} = 1\, ,
\ee
 to extract
the 
renormalization factors $Z_q$ and $Z_{\cal O}$. These conditions are
imposed in the massless theory and 
therefore a chiral extrapolation is needed. The mass-dependence is very weak
for the vector and axial vector operators. This is demonstrated 
 in Fig.~6  for the case of the one-derivative  vector the axial-vector operators using $N_f=2$ TMF~\cite{Z-Alexandrou}. Results on 
 $Z_V$ and $Z_A$ as a function of $(ap)^2$
 are shown in Fig.~7  where plateaux are improved after 
subtracting ${\cal O}(a^2)$-terms perturbatively~\cite{Z2-Alexandrou}. Using the  RI$^\prime$-MOM  scheme but with a momentum independent source,
the RBC-UKQCD Collaborations made a comparison between perturbative  and non-perturbative determination of the renormalization constants
and found  that results for $\langle x \rangle_{u-d} $~\cite{Aoki} with perturbative renormalization are lower bringing them in
agreement with LHPC's results, which used perturbative renormalization~\cite{FF-Bratt}. It is therefore important 
to compute the renormalization constants non-perturbatively.

      \begin{minipage}{0.5\linewidth}
 \hspace*{-2cm}     \includegraphics[width=1.5\linewidth]{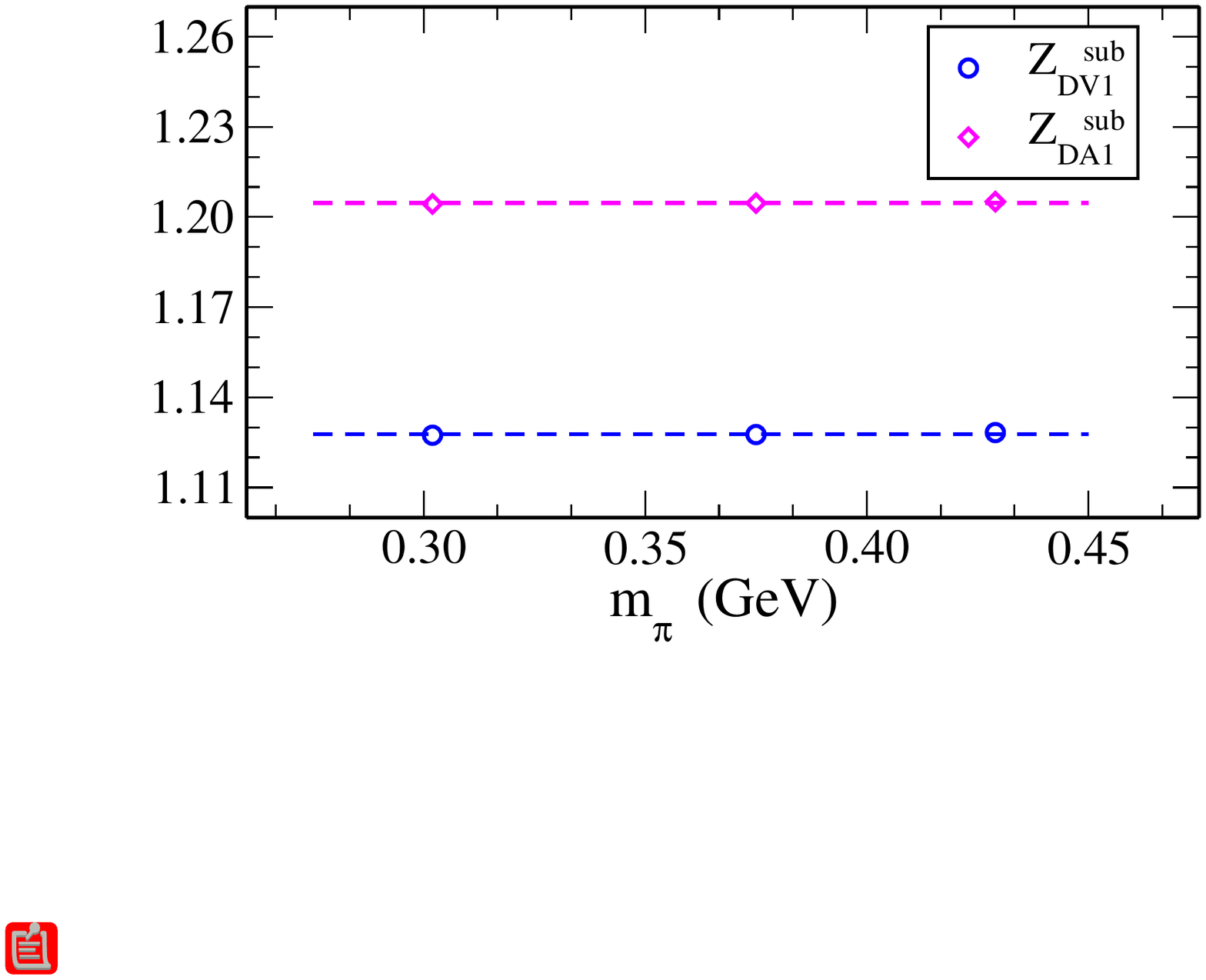}\vspace*{-2.8cm}\\
{\hspace*{-0.8cm}\small {\bf Figure 6:} Mass dependence of the renormalization \\
\hspace*{-0.8cm} constant for
vector and axial vector one-derivative\\ \hspace*{-0.8cm}operators
for $N_f=2$ TMF~\cite{Z-Alexandrou}.} \\
\end{minipage}\hfill
      \begin{minipage}{0.4\linewidth}
 \hspace*{-1.2cm}{\includegraphics[width=1.3\linewidth, height=1.6\linewidth]{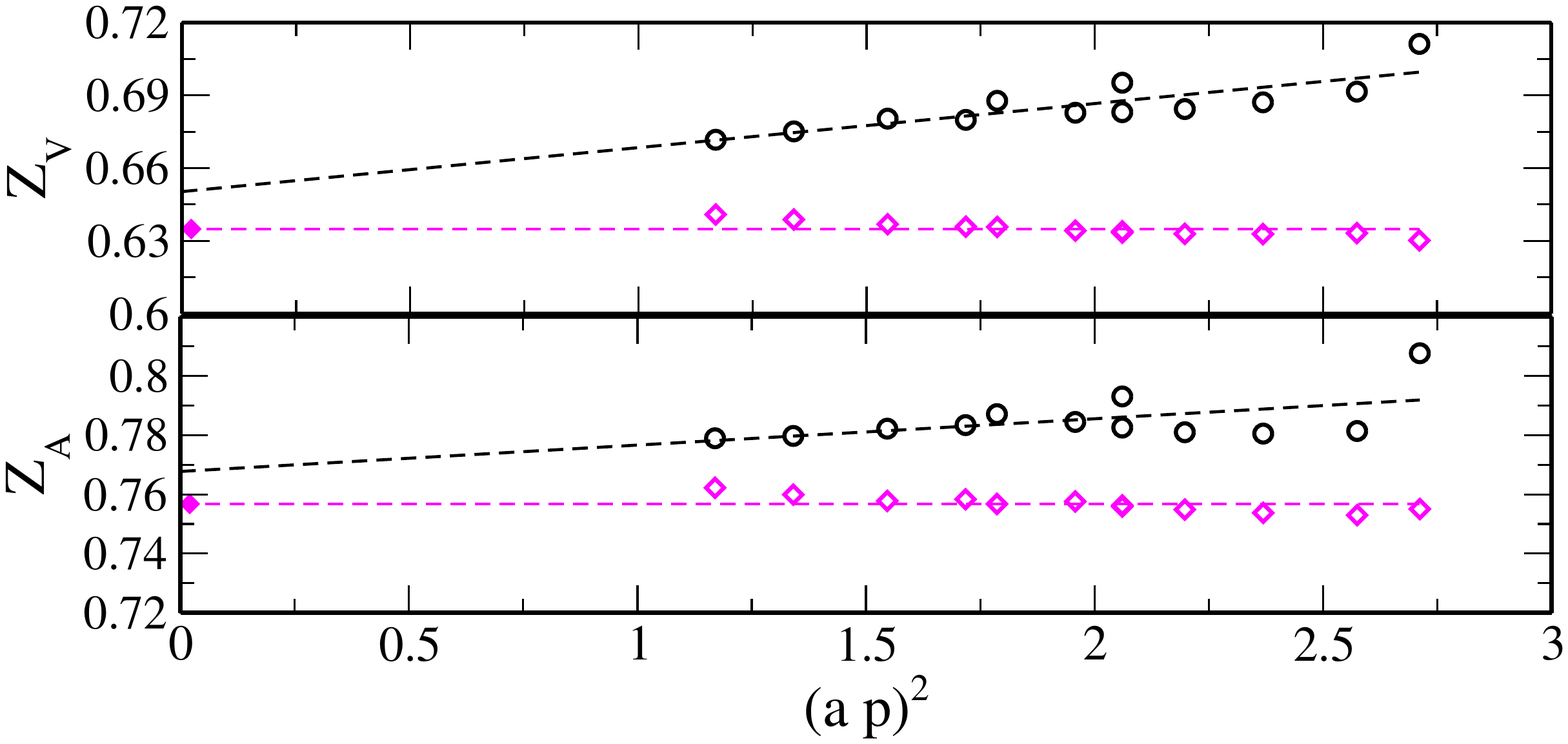}}\vspace*{-3cm}\vspace*{-1cm}\\
{\small \hspace*{-1cm}{\bf Figure 7:} $Z_V$ and $Z_A$ with perturbative subtraction \\
\hspace*{-1cm} of  ${\cal O}(a^2)$-terms  for $N_f=2$ TMF~\cite{Z2-Alexandrou}.}
\end{minipage}

 $\bullet$ {\bf Cut-off effects:}
 The  nucleon axial charge $g_A$, the isovector momentum fraction $<x>_{u-d}=A_{20}(0)$ and helicity fraction $\langle x\rangle_{\Delta u-\Delta d}=\tilde{A}_{20}(0)$ are calculated directly at $Q^2=0$ requiring no fits.  We can examine their dependence on the lattice spacing
by obtaining these quantities at a given value of the pseudoscalar mass
in units of $r_0$. In Fig.~8 we show results at three lattice spacings using $N_f=2$ TMF. As can be seen, ${\cal O}(a^2)$-terms are small and, 
allowing a linear dependence, yields consistent results to those obtained with a constant fit.   This is
also true for the nucleon isovector anomalous moment $\kappa_v$, Dirac and Pauli radii $r_1^2$ and $r_2^2$ that require fits to the EM form factors.

\begin{minipage}{0.49\linewidth}
   \includegraphics[width=\linewidth]{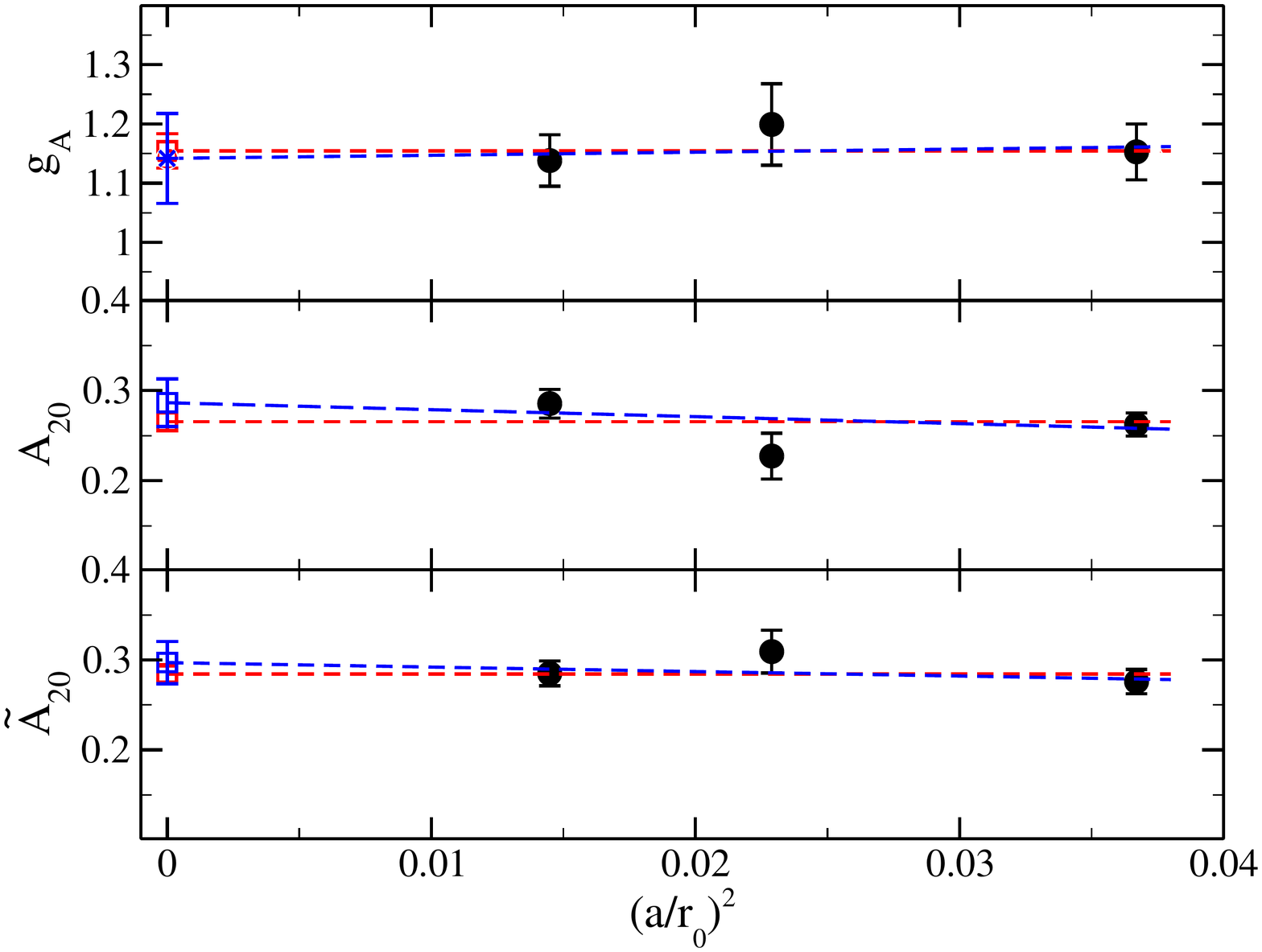}
\end{minipage}\hfill
\begin{minipage}{0.49\linewidth}
 {  \includegraphics[width=\linewidth]{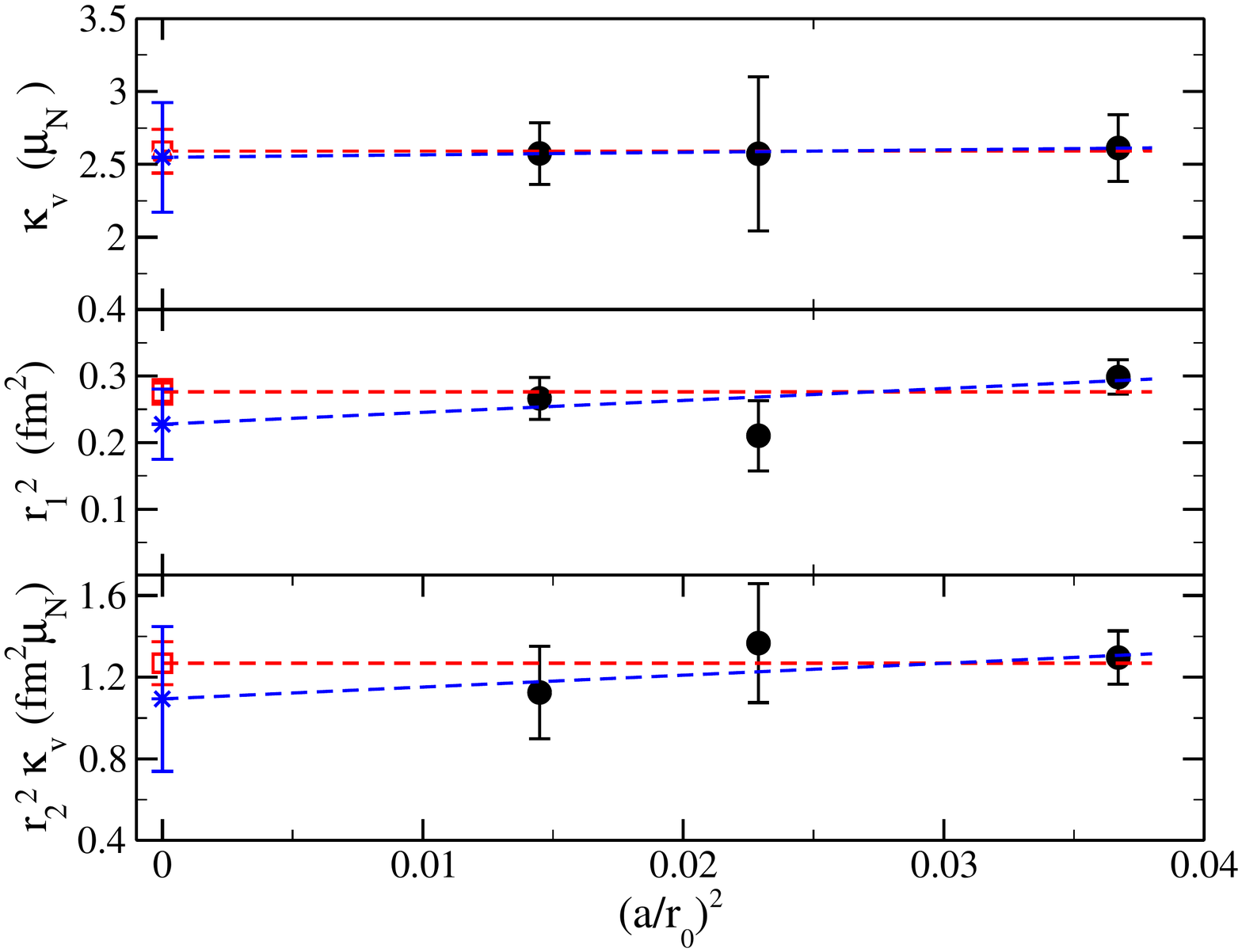}}
\end{minipage}
{\small {\bf Figure 8:} Left: $g_A$, $\langle x \rangle_{u-d}$, and $\langle x \rangle_{\Delta u-\Delta d}$; Right: the nucleon anomalous magnetic moment $\kappa_v$, the Dirac radius $r_1^2$ and Pauli radius $r_2^2$ times $\kappa_v$ as a function of $(a/r_0)^2$. The red line is the result of fitting to a constant; the blue one is a linear fit. The results are obtained using $N_f=2$ TMF~\cite{FF-Alexandrou}.}    

\noindent
We therefore conclude that cut-off effects are small for $a<0.1$~fm  for ${\cal O}(a)$-improved actions and that one can use continuum chiral perturbation theory  to extrapolate to the physical limit.

 $\bullet$ {\bf Finite volume corrections:}
In Fig.~9 we compare results on  $g_A$,  $\langle x \rangle _{u-d}$ and $\langle  x \rangle _{\Delta u-\Delta d}$ computed  on different lattice sizes 
 as a function of $m_\pi^2$.

\begin{minipage}{0.49\linewidth}\vspace*{-1.cm}
 { \hspace*{-2.cm}\includegraphics[width=1.3\linewidth]{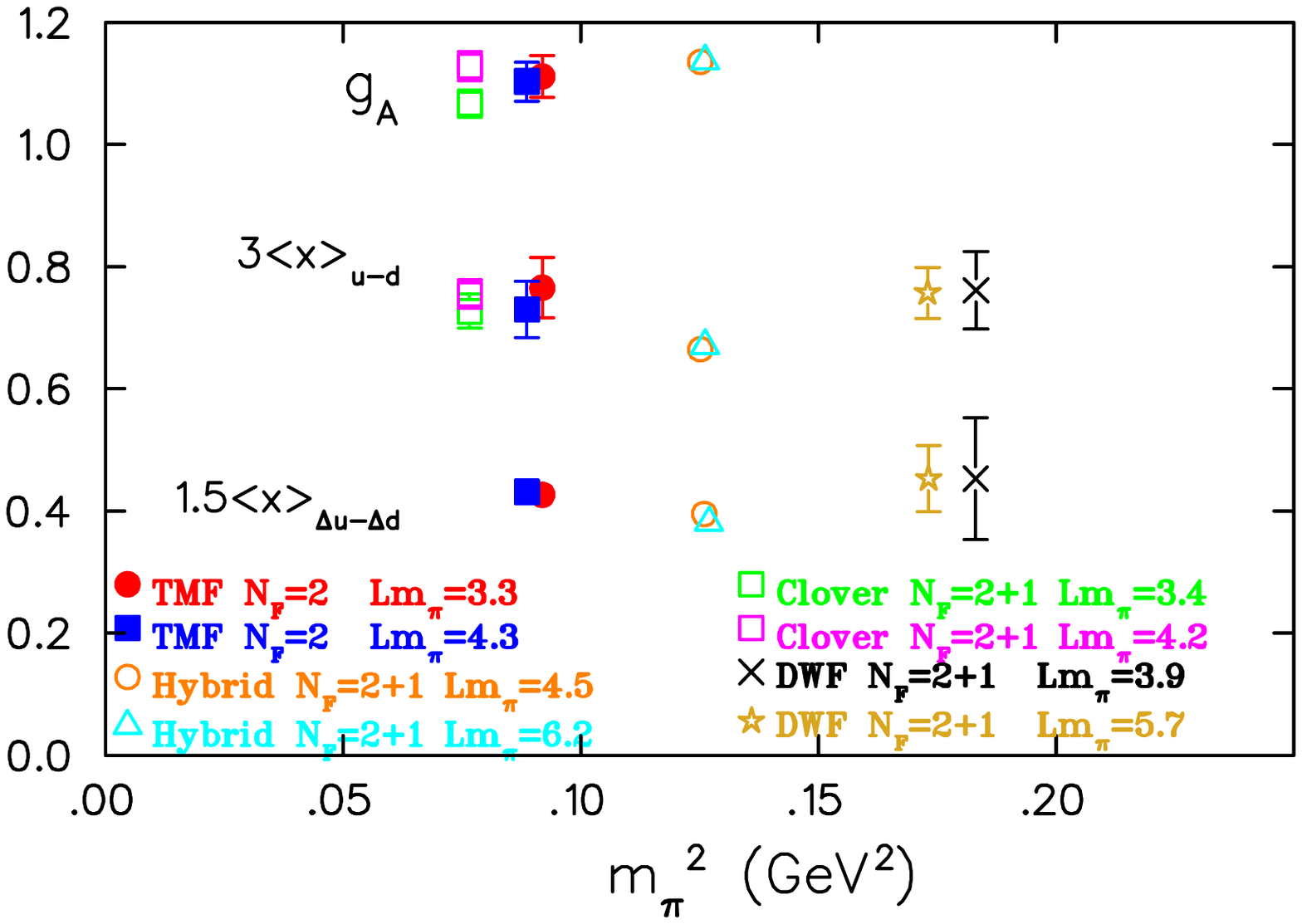}}
\vspace*{-6.5cm}\\{\small \hspace*{-0.8cm}{\bf Figure 9:}   $g_A$, 3$\langle x\rangle_{u-d}$ and $\frac{3}{2}\langle x\rangle_{\Delta u-\Delta d}$ using TMF~\cite{FF-Alexandrou}, \\
\hspace*{-0.8cm}Clover~\cite{GPD-Zanotti}, hybrid~\cite{FF-Bratt} and DWF~\cite{Aoki}.}
\end{minipage}\hfill
\begin{minipage}{0.49\linewidth}
The observations are:  i) Accurate lattice data by LHPC using 
 domain wall valence quarks on staggered sea (hybrid) for $m_\pi\sim 350$~MeV  with $Lm_\pi=4.5$ and  $Lm_\pi=6.2$
show no  volume effects;
ii)  TMF results for $m_\pi\sim 300$~MeV with 
 $Lm_\pi=3.3$ and  $Lm_\pi=4.3$ are consistent;
iii)  Results for $\langle x \rangle _{u-d}$ using Clover fermions from  QCDSF  for $m_\pi \sim 270$~MeV 
with $Lm_\pi=3.4$ and $Lm_\pi=4.2$ are consistent, whereas 
 $g_A$  differs by about a standard deviation;
iv)  RBC-UKQCD results with domain wall fermions (DWF) with $Lm_\pi=3.9$ and $Lm_\pi=5.7$ show no  volume
effects for $\langle x \rangle _{u-d}$ and $\langle x\rangle_{\Delta u- \Delta d}$
~\cite{Aoki}.

\end{minipage}

\vspace*{0.3cm}

\noindent
Within the current statistical uncertainties this comparison, therefore  shows that volume effects on  $<x>_{u-d}$   and $<x>_{\Delta u-\Delta d}$ are negligible 
 for $Lm_\pi\stackrel{>}{\sim} 3.3$. 
 For $g_A$ volume effects of about one standard deviation
 are seen for $Lm_\pi{\sim} 3.4$ and 
pion mass below 300~MeV and therefore  larger lattices keeping   $Lm_\pi\stackrel{>}{\sim} 4$ may be needed.

Comparing the isovector  nucleon EM FFs as a function of $Q^2$
at $m_\pi\sim 300$~MeV  we also find consistent results
for $Lm_\pi=3.3$ and $Lm_\pi=4.3$.
A similar behavior is also observed for the nucleon axial FF $G_A(Q^2)$
 whereas the induced pseudoscalar FF $G_p(Q^2)$, which has a pion pole 
behavior, may suffer from larger finite volume corrections at low $Q^2$-values~\cite{Alexandrou-new}.



\section{Results on nucleon form factors}
Having examined cut-off and volume effects we compare results from various
collaborations computed with dynamical fermions for $a\stackrel{<}{\sim} 0.1$~fm\footnote{We note that results by the LHPC using a hybrid action have $a=0.124$~fm.} and $Lm_\pi \stackrel{>}{\sim}3.3$.

$\bullet$ {\bf Nucleon axial charge:}
The axial charge is well known experimentally. Since it is determined at $Q^2=0$
there is no ambiguity associated with fitting the $Q^2$-dependence
 of the FF. In Fig.~\ref{fig:gA} we show  recent
 lattice results using TMF, DWF and a hybrid action of DWF
on a staggered sea, all of which are renormalized non-perturbatively.
As can be seen, there is a nice 
 agreement among different 
lattice discretizations and no significant dependence on the quark mass down
to about $m_\pi=260$~MeV.

\addtocounter{figure}{5}

\begin{figure}[h]\vspace*{-4cm}
\begin{minipage}{0.55\linewidth}
  { \hspace*{-1cm}\includegraphics[width=1.1\linewidth]{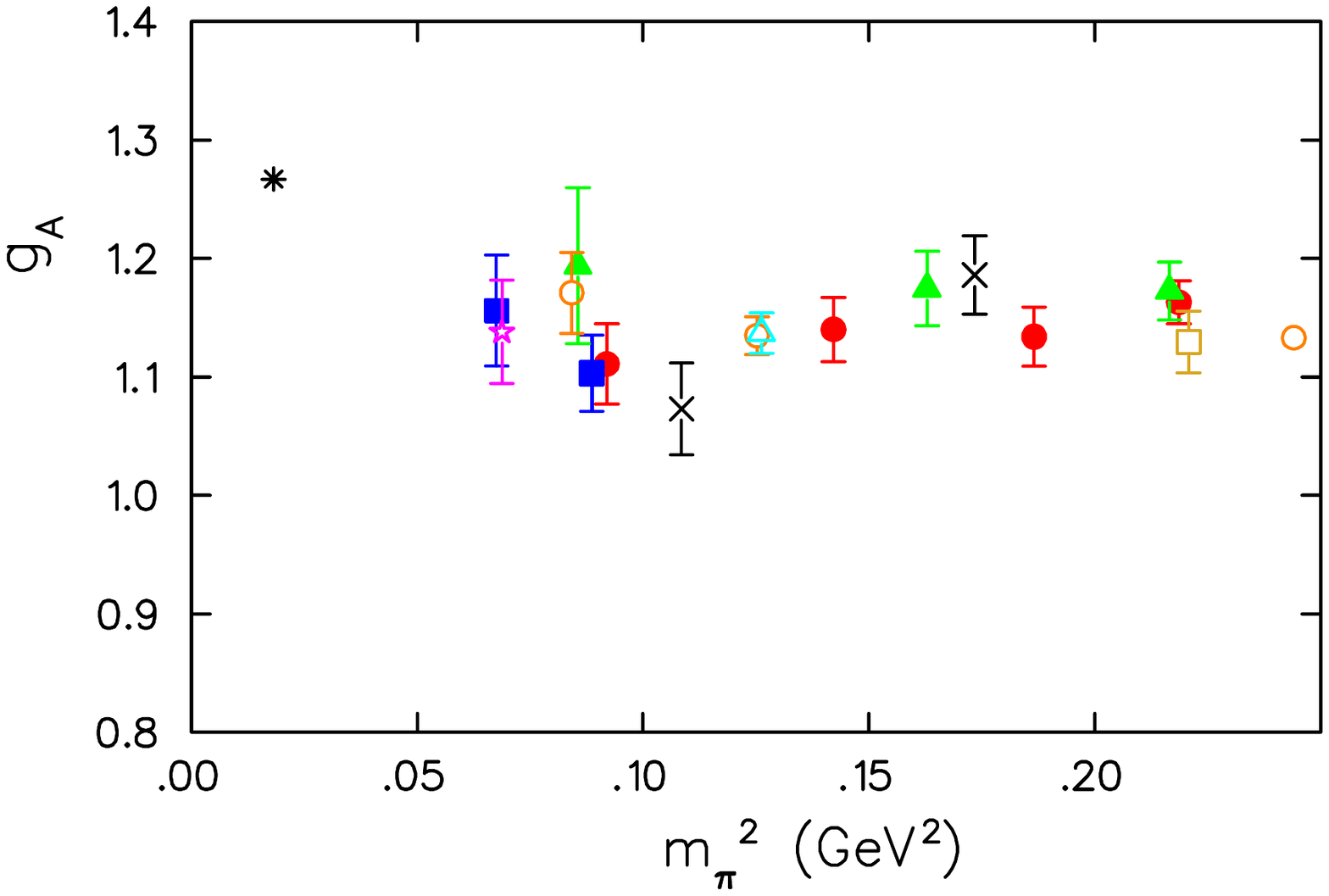}}
\end{minipage}\hfill
\begin{minipage}{0.55\linewidth}
  { \hspace*{-1cm}\includegraphics[width=1.1\linewidth]{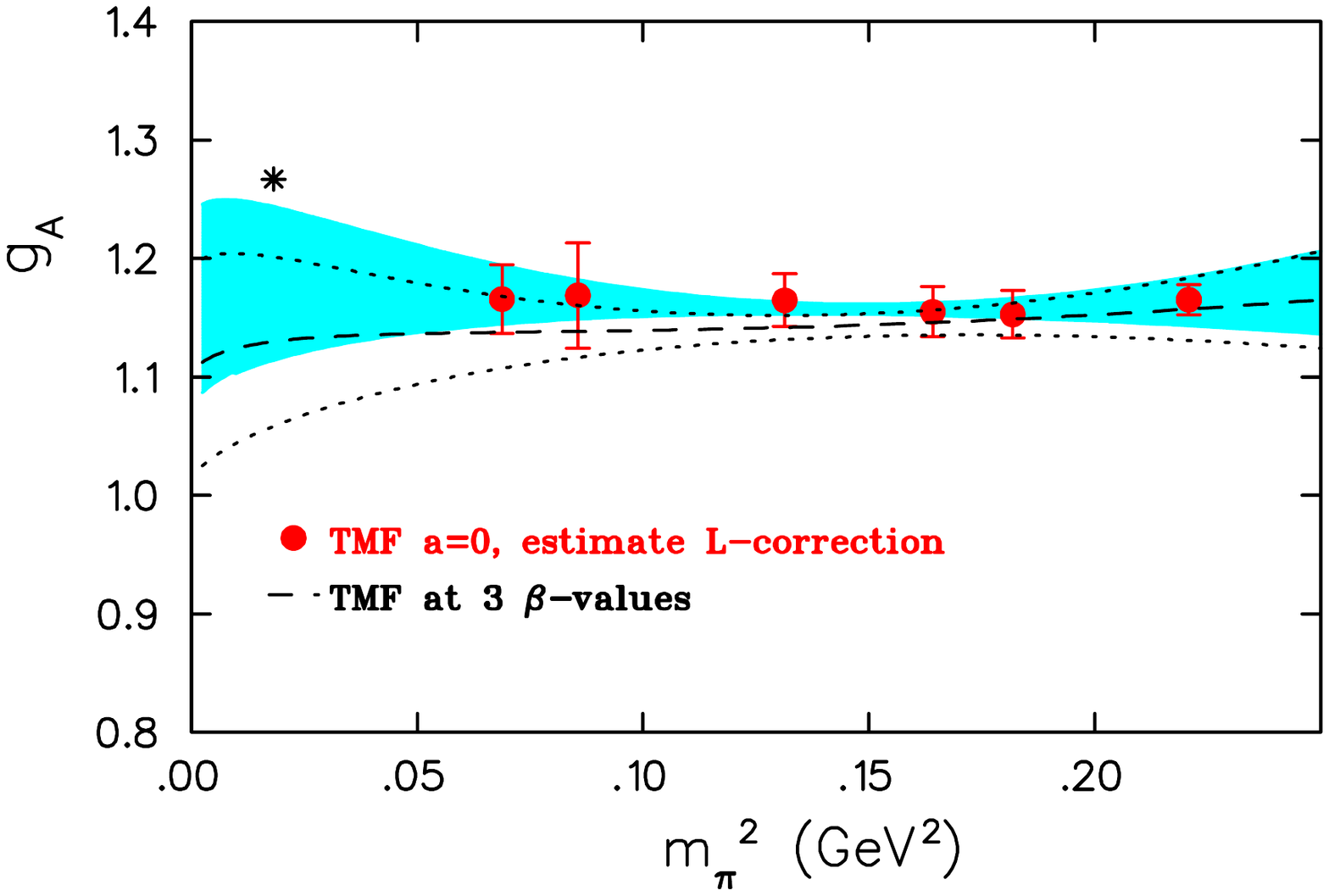}}
\end{minipage}\vspace*{-2.5cm}
\caption{Left: Lattice data on $g_A$ using $N_f=2$ TMF~\cite{FF-Alexandrou} 
($a=0.089$~fm: filled red
circles for $L=2.1$~fm and filled blue squares for $L=2.8$~fm;  $a=0.070$~fm: filled green triangles for $L=2.2$~fm; $a=0.056$~fm:
purple star for $L=2.7$~fm and open yellow square for $L=1.8$~fm),  $N_f=2+1$  DWF~\cite{FF-Yamazaki} (crosses for $a=0.114$~fm and $L=2.7$~fm) and $N_f=2+1$ using DWF and staggered sea~\cite{FF-Bratt} ($a=0.124$~fm: open orange circles for $L=2.5$~fm and open cyan triangle for $L=3.5$~fm). The physical point is shown by the asterisk.
Right: Volume corrected TMF results extrapolated to the continuum limit  
 together with the fit  using HB$\chi$PT (blue band). The band bounded by the lines is the resulting fit to the TMF data shown on the left.} 
\label{fig:gA}
\end{figure}

To illustrate the size of lattice artifacts and obtain a value of $g_A$
 at the physical point, we use TMF results~\cite{Alexandrou-new}.
The volume corrected~\cite{Khan} data are extrapolated
to $a=0$  using 
 three
lattice spacings, namely $a=0.089$~fm, $0.070$~fm  and $0.056$~fm,  at
two values of the pseudoscalar mass, by fitting to a
constant. For intermediate masses we use data at the two coarser lattices.
The  continuum volume-corrected results are shown in Fig.~\ref{fig:gA}. Chiral extrapolation
using one-loop heavy baryon chiral perturbation theory (HB$\chi$PT) in the small scale expansion (SSE)~\cite{SSE}
with three fit parameters   produces a value of $g_A=1.12(8)$ at the
physical point, which is lower than the experimental value
 by about a standard deviation. The large
error is due to the strong correlation between the $\Delta$ axial
charge $g_{\Delta \Delta}$ and the counter-term  involved in the fit.
 Therefore, a lattice determination of $g_{\Delta \Delta}$ will allow a more controlled
chiral extrapolation. Fitting
the raw lattice data produces the band shown by the dotted lines. Therefore
one observes that, although the continuum volume-corrected results
are closer to experiment, the largest uncertainty is due to
the chiral extrapolation and at the physical point 
the values obtained using the raw and continuum volume-corrected lattice data
are consistent.

\begin{figure}[h]\vspace*{-3.5cm}
\begin{minipage}{0.49\linewidth}
      {\includegraphics[width=1.1\linewidth]{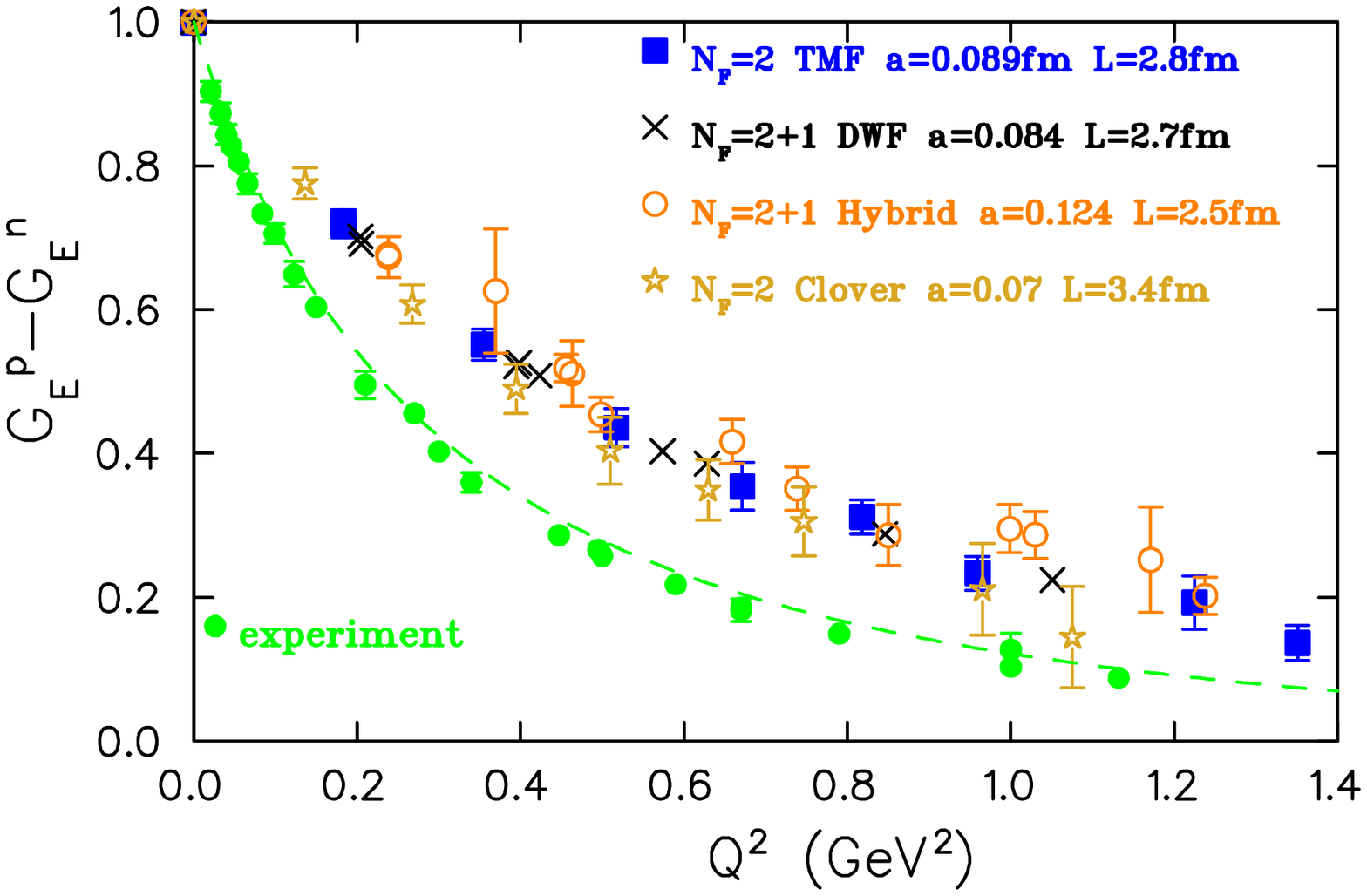}}\vspace*{-5.5cm}
      {\includegraphics[width=1.1\linewidth]{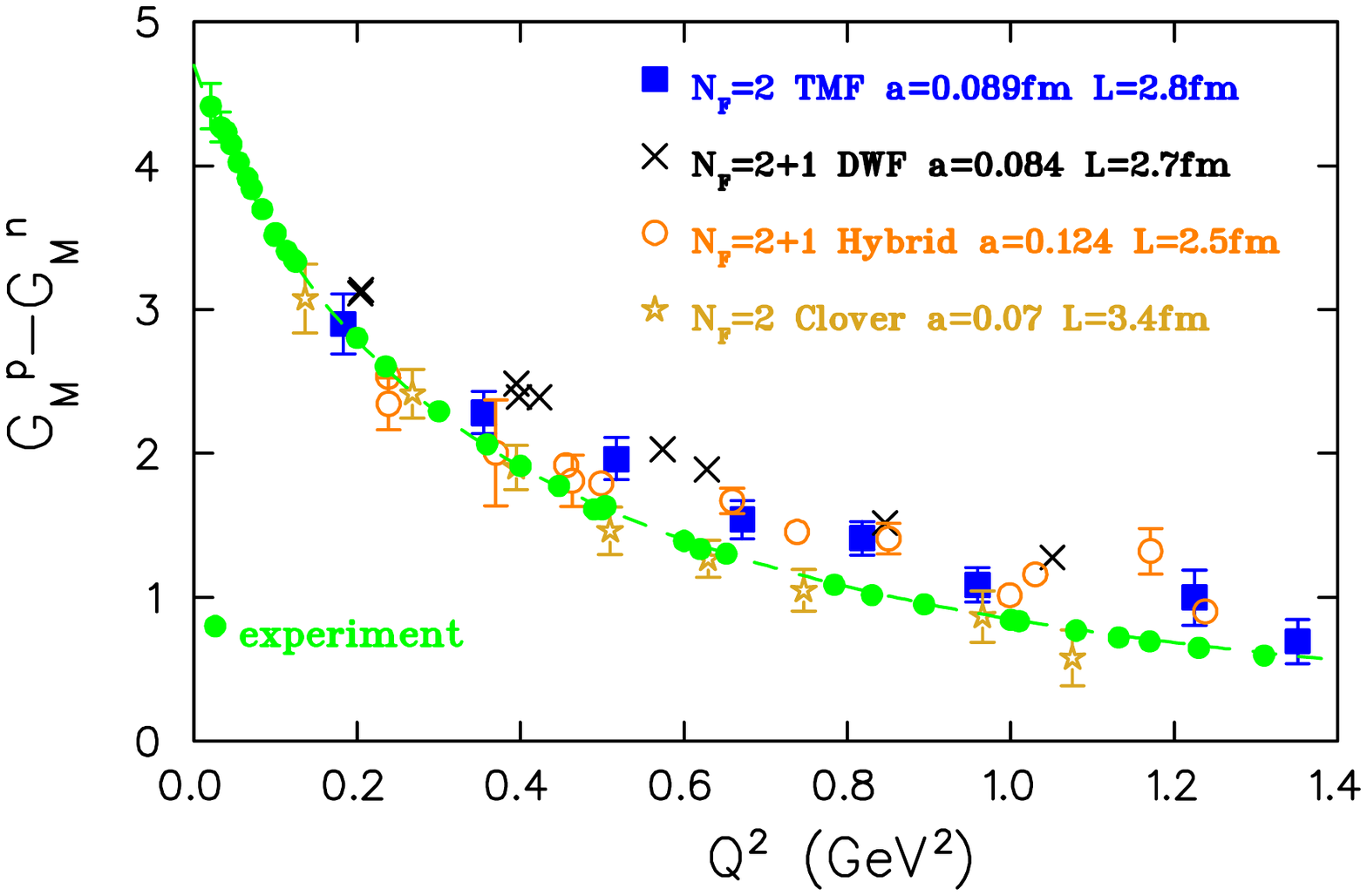}}
   \end{minipage}
\begin{minipage}{0.49\linewidth}
   {\includegraphics[width=1.1\linewidth]{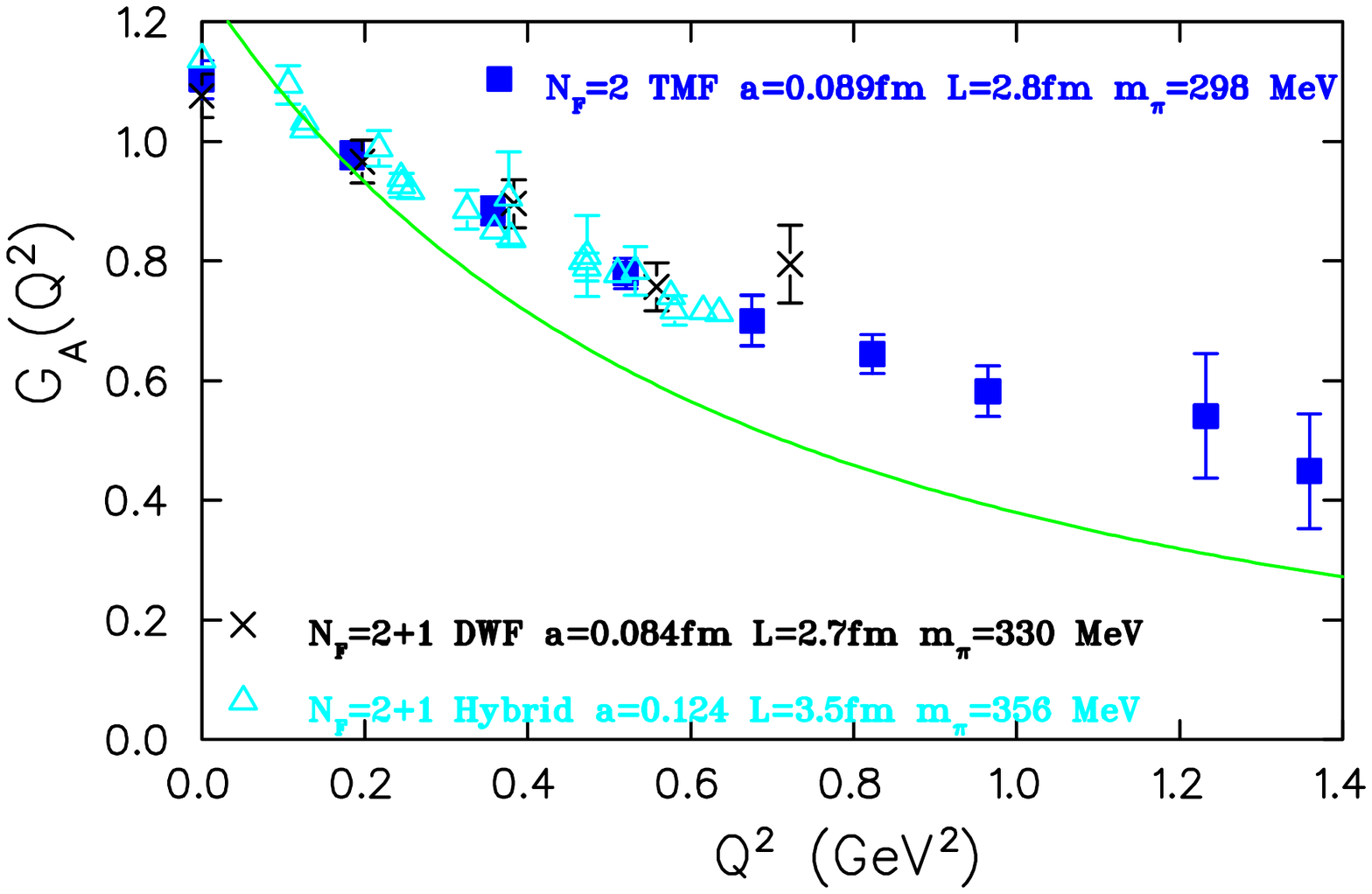}}\vspace*{-5.8cm}
     {\includegraphics[width=1.1\linewidth]{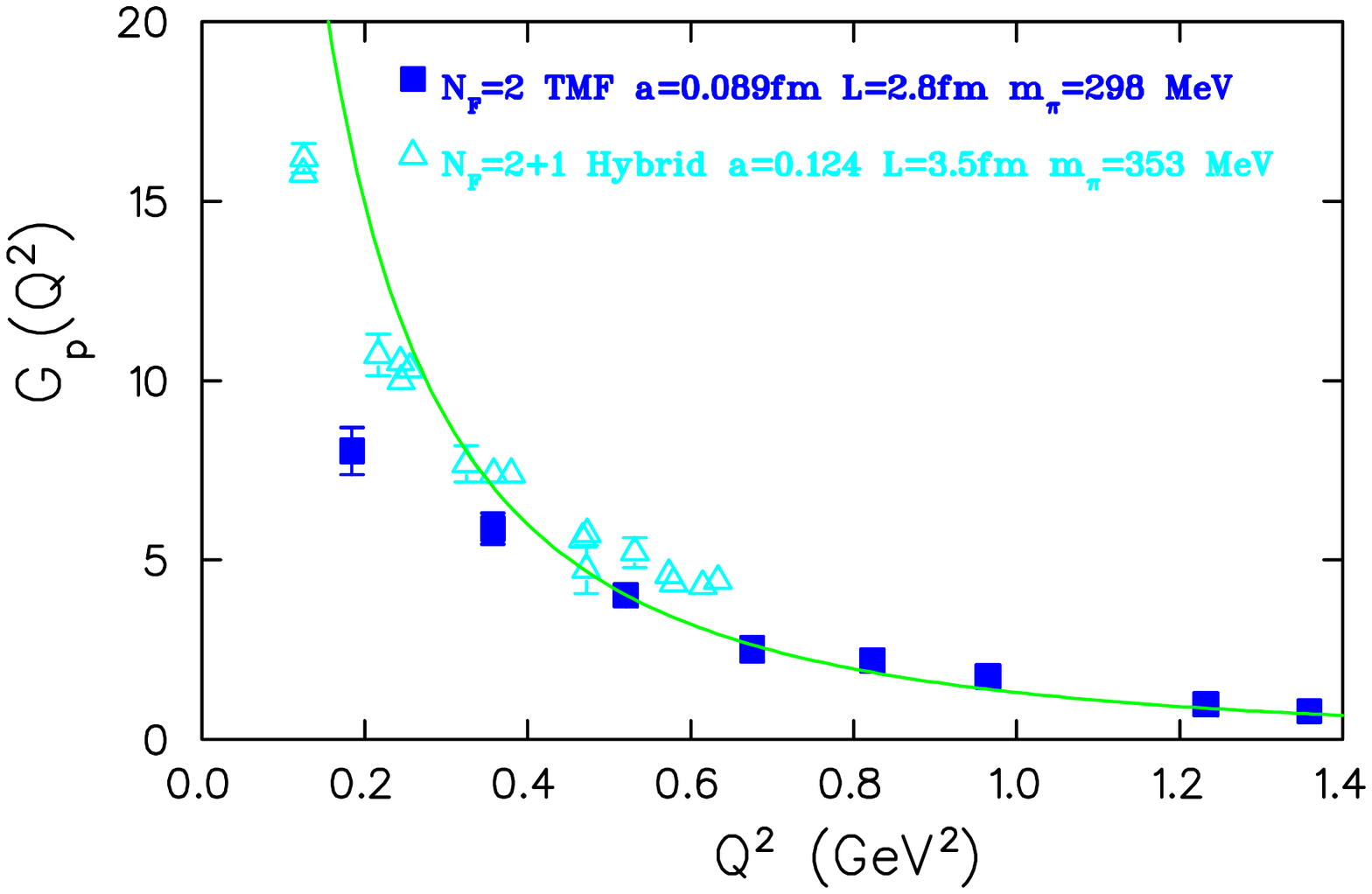}}
   \end{minipage}\vspace*{-2.cm}
\caption{Left: Isovector electric and magnetic nucleon FFs at $m_\pi\sim 300$~MeV using TMF~\cite{FF-Alexandrou} (filled  blue squares) DWF~\cite{FF-Syritsyn}
 (crosses), hybrid~\cite{FF-Bratt} (open orange circles) and Clover fermions~\cite{Wittig} (yellow stars). Experimental data are shown with the filled green circles 
accompanied with Kelly's parametrization shown with the dashed line.  
Right: Axial nucleon FFs. The solid line is a dipole fit to experimental data for  $G_A(Q^2)$ combined with pion pole dominance to
get the solid curve shown for $G_p(Q^2)$.}
\label{fig:nucleon EM}
\end{figure}

$\bullet$ {\bf Nucleon form factors:}
Recent lattice results on the EM isovector and axial FFs   are shown in Fig.~\ref{fig:nucleon EM}. We observe
a nice agreement among lattice results, in particular  for $G_E(Q^2)$ and $G_A(Q^2)$. However, both $G_E(Q^2)$ and $G_A(Q^2)$  decrease with $Q^2$ less rapidly than 
experiment. We note that
a good description of the
$Q^2-$ dependence for both $G_E(Q^2)$ and $G_M(Q^2)$ is provided by a dipole
 form 
using the lattice-computed $\rho-$meson mass.
 Lattice results on $G_p(Q^2)$ using TMF
and those 
obtained using the hybrid action on a larger volume, are not
consistent, in particular at small $Q^2$
where $G_p(Q^2)$  increases rapidly
 due to the pion-pole behavior. From the observed quark mass dependence of $G_p(Q^2)$~\cite{Alexandrou-new}  the 50~MeV difference in the pion mass may not be sufficient to 
fully account for this discrepancy, which may indicate  volume effects.

\begin{figure}[h]\vspace*{-0.5cm}
\begin{minipage}{0.5\linewidth}\vspace*{-0.5cm}
   {\includegraphics[width=\linewidth]{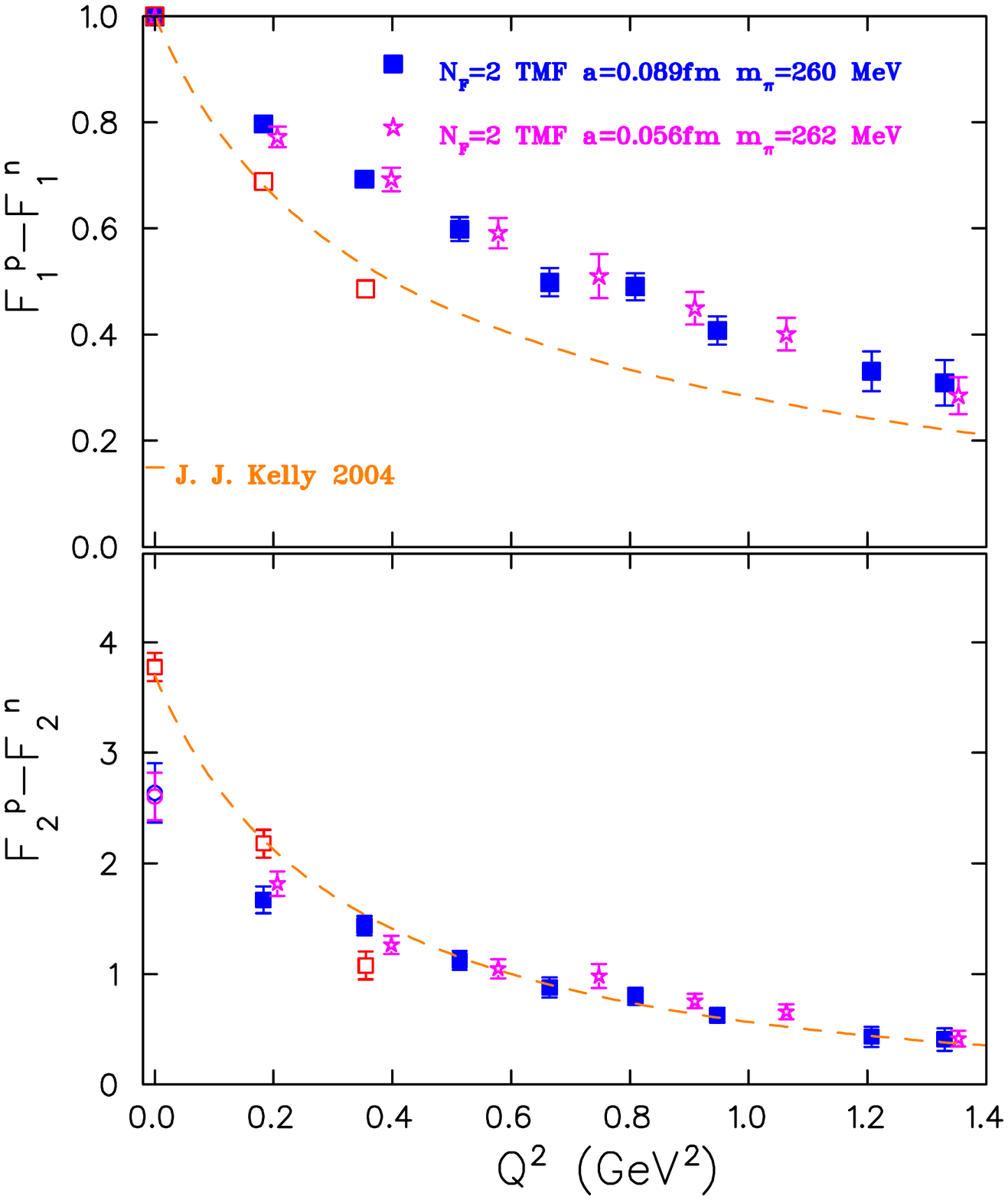}}
\end{minipage}
\begin{minipage}{0.49\linewidth} \vspace*{-0.5cm}
{\includegraphics[width=\linewidth]{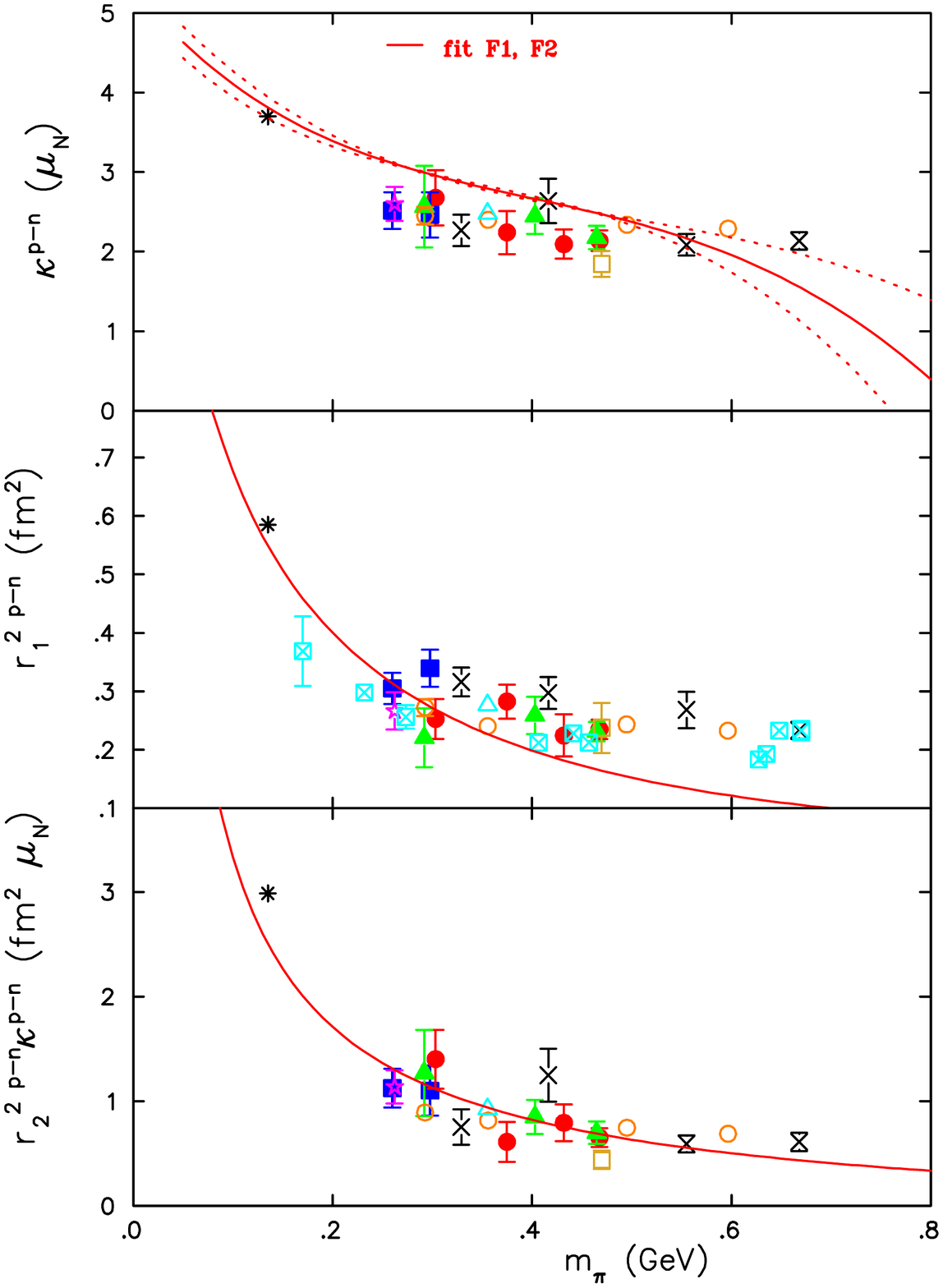}}
\end{minipage}
\vspace*{-0.5cm}\caption{Left: Open squares show the chirally extrapolated results at the physical point. The
dashed line is Kelly's parametrization of the experimental data.
Right: The solid lines show the prediction of HB$\chi$PT using the
parameters obtained from fitting $F_1(m_\pi, Q^2)$ and $F_2(m_\pi,Q^2)$. The
notation is the same as that of Fig.~10. For $r_1^2$ results 
using Clover fermions~\cite{GPD-Zanotti} are also shown with the cyan cross-in-square symbols.}
\label{fig:kappav}
\end{figure}

Using HB$\chi$PT to one-loop, with $\Delta$ degrees
 of freedom and iso-vector  $N$-$\Delta$ coupling included in LO~\cite{FF-Hemmert}
we perform a  fit  to 
$F_1(m_\pi,Q^2)$ and $F_2(m_\pi,Q^2)$ with five parameters, namely the
iso-vector magnetic moment at the chiral limit {$\kappa^0_v$}, 
the isovector and axial N to $\Delta$ 
coupling constants and two counterterms.
As can be seen, the chiral extrapolation increases the value of $F_1$ and $F_2$
at low $Q^2$,  bringing it into qualitative agreement with experiment.
Application of twisted b.c. with
a study of the associated volume corrections~\cite{Hagler} will
be very useful in enabling us to
obtain these FFs at lower $Q^2$-values,  permitting a better chiral extrapolation.
 Using the parameters extracted from the fits to $F_1$ and $F_2$ we
obtain the chiral dependence of the isovector nucleon anomalous magnetic moment
and the Dirac and Pauli radii shown in Fig.~\ref{fig:kappav}.

\vspace*{-0.3cm}

\section{Results on nucleon moments}
In this section we show results on the nucleon matrix
element of the one-derivative operators
 $\ \ \bar u \gamma_{\{\mu} \stackrel{\leftrightarrow}{ D}_{\nu\}} u - \bar d \gamma_{\{\mu} \stackrel{\leftrightarrow}{ D}_{\nu\}} d$  and 
  $\ \ \bar u \gamma_5\gamma_{\{\mu} \stackrel{\leftrightarrow}{ D}_{\nu\}} u - \bar d \gamma_5\gamma_{\{\mu} \stackrel{\leftrightarrow}{ D}_{\nu\}} d$ in the  $\overline{MS}$ scheme at a scale $ \mu=2$~GeV.

\begin{figure}[h]\vspace*{-0.2cm}
\begin{minipage}{0.48\linewidth}
{\includegraphics[width=\linewidth]{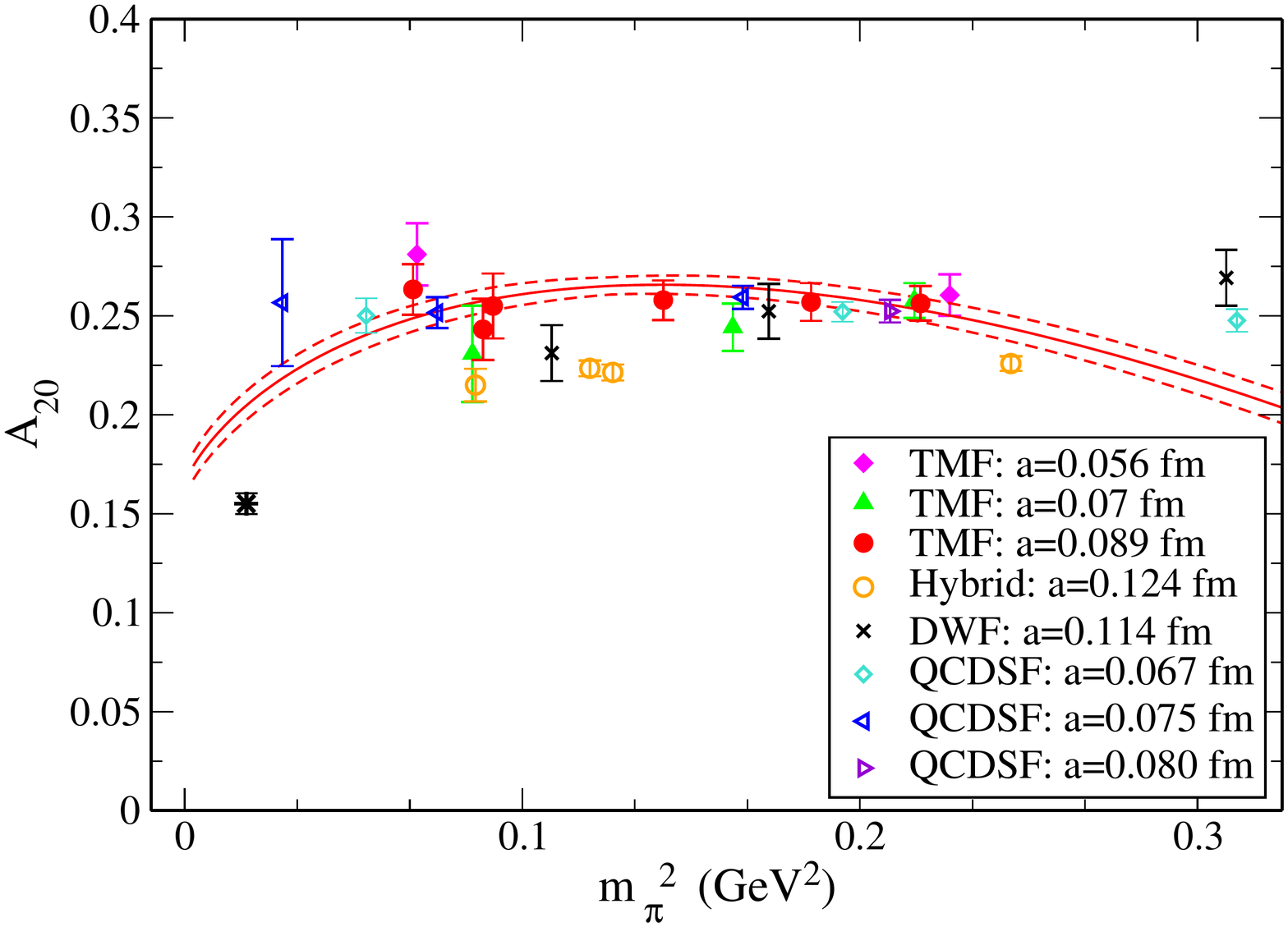}}
\end{minipage}\hfill
\begin{minipage}{0.48\linewidth}
{\includegraphics[width=\linewidth]{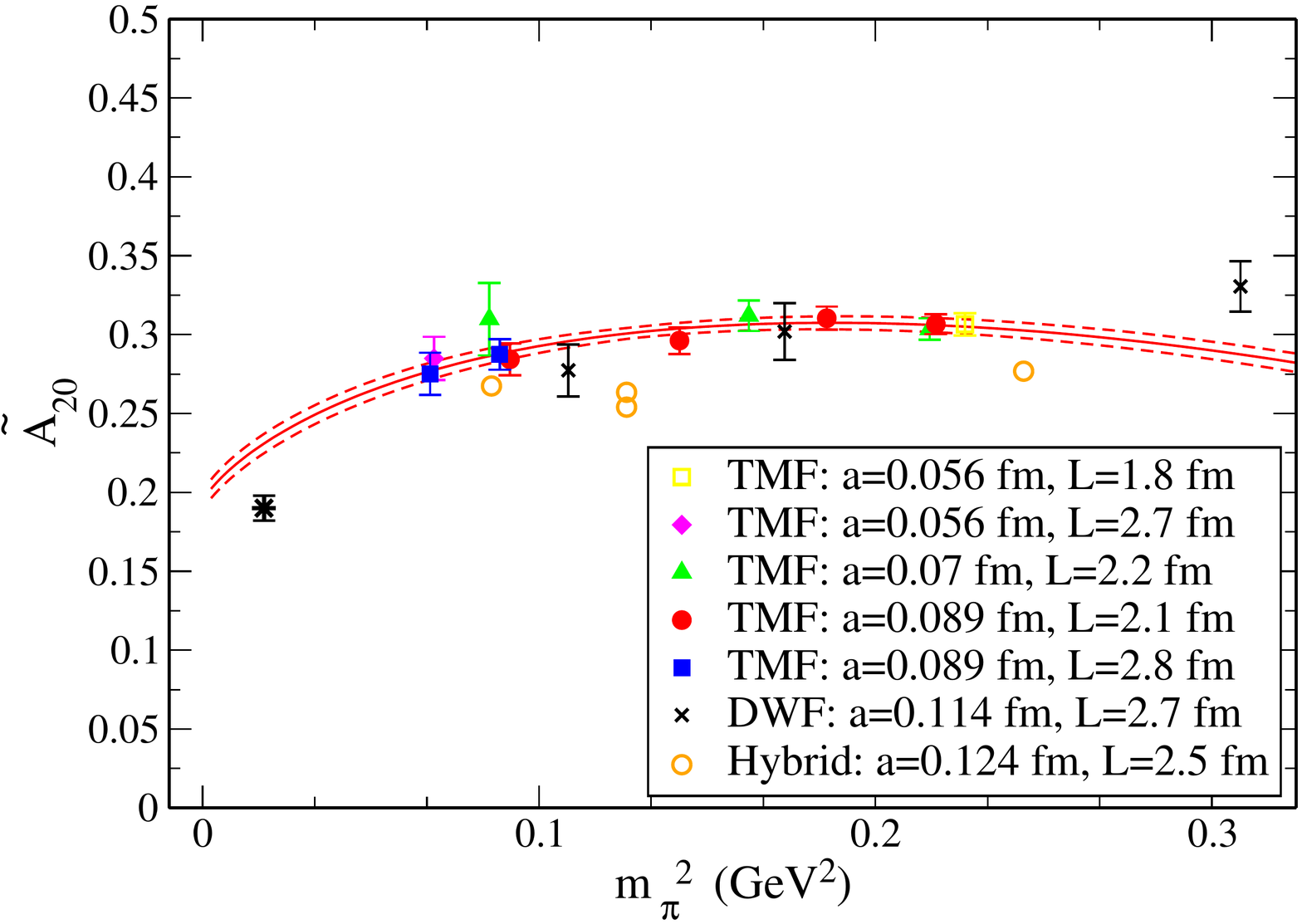}}
\end{minipage}
\caption{Recent results on the isovector $A_{20}= \langle x \rangle _{u-d}$ and
 $\tilde{A}_{20}=\langle x \rangle _{\Delta u-\Delta d}$. } 
\label{fig:GPDs}
\end{figure}


In Fig.~\ref{fig:GPDs} we compare recent results from ETMC~\cite{korzec}, RBC-UKQCD~\cite{Aoki}, QCDSF~\cite{GPD-Zanotti} and LHPC~\cite{FF-Bratt} on
the spin-independent and helicity quark distributions.
All collaborations except LHPC use non-perturbatively computed renormalization constants. As already mentioned, The ETMC has, in addition, subtracted ${\cal O}(a^2)$ terms perturbatively to reduce lattice artifacts~\cite{Z-Alexandrou}.
There is a spread in the values of the lattice results. It was noted that taking
a renormalization free ratio leads to a better agreement among lattice data with
$Lm_\pi>4$~\cite{Dru}.  In particular this brought the LHPC data in agreement
with those from ETMC and QCDSF.



In HB$\chi$PT~\cite{Thomas} the expressions for the $m_\pi$-dependence of 
$A_{20}$ and $\tilde{A}_{20}$ are given by:

$$ 
{\scriptsize
\langle x \rangle_{u-d} = {C}\left[1-\frac{3g_A^2+1}{(4\pi f_\pi)^2} m_\pi^2 \ln\frac{m_\pi^2}{\lambda^2} \right] + \frac{{c_8}(\lambda^2) m_\pi^2}{(4\pi f_\pi)^2}\,, \hspace*{0.5cm}
\langle x\rangle_{\Delta u-\Delta d} ={\tilde{C}}\left[1-\frac{2g_A^2+1}{(4\pi f_\pi)^2} m_\pi^2 \ln\frac{m_\pi^2}{\lambda^2} \right]+ \frac{{\tilde{c}_8}(\lambda^2) m_\pi^2}{(4\pi f_\pi)^2}  \nonumber }
$$
Using $\lambda^2=1$~GeV$^2$ and the TMF results we obtain the curves shown in 
Fig.~\ref{fig:GPDs},
which yield a value higher than experiment for both observables.
The very recent result by QCDSF~\cite{GPD-Zanotti} at $m_\pi\sim 170$~MeV 
 remains higher than experiment and  highlights the need to understand such deviations.

\section{$N$-$\Delta$ system}
\begin{figure}[h]\vspace*{-0.8cm}
{\begin{minipage}{0.33\linewidth}\vspace*{-4.cm}
\hspace*{-0.5cm}{\includegraphics[width=1.12\linewidth,height=0.76\linewidth]{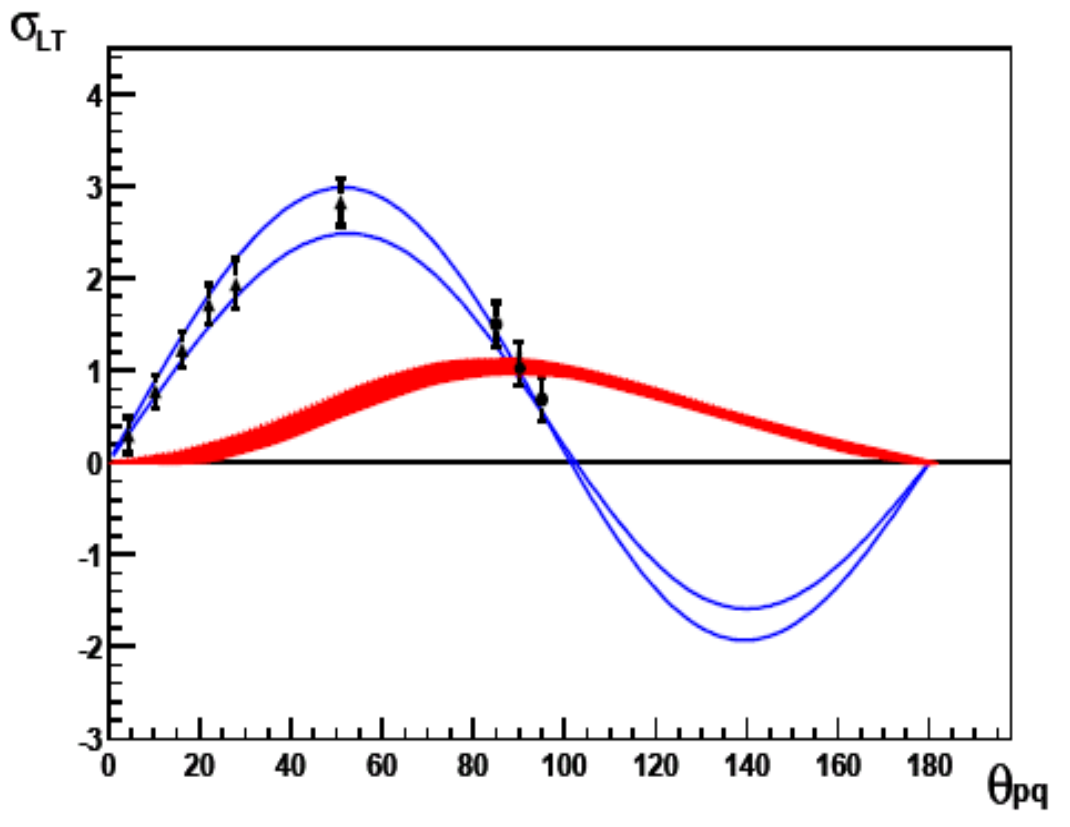}}
\end{minipage}\hfill
\begin{minipage}{0.33\linewidth}
\hspace*{-0.7cm}{\includegraphics[width=1.35\linewidth,height=1.8\linewidth]{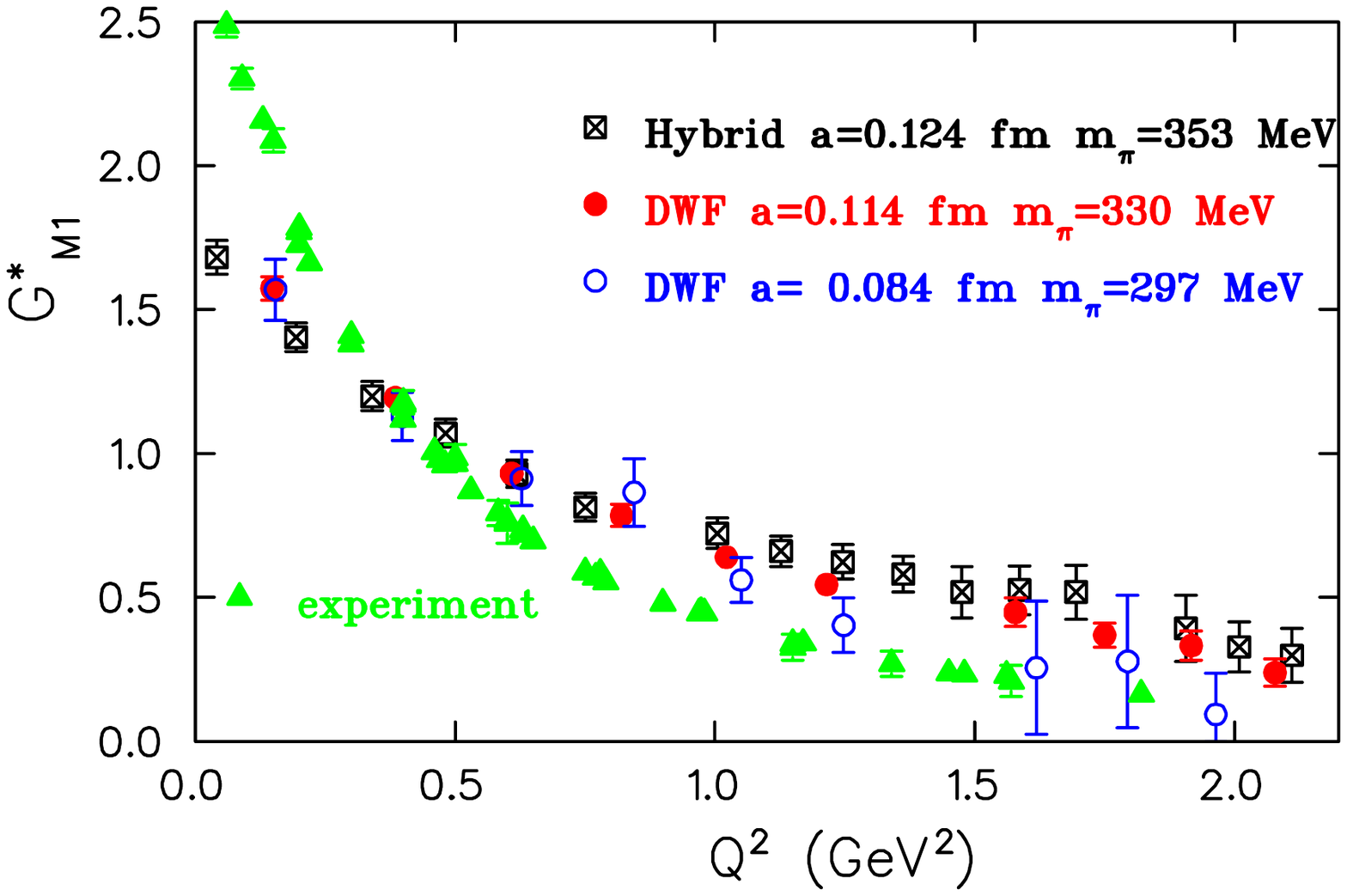}}
\end{minipage}\hfill
\begin{minipage}{0.33\linewidth}
\hspace*{-0.2cm}{\includegraphics[width=1.35\linewidth,height=1.8\linewidth]{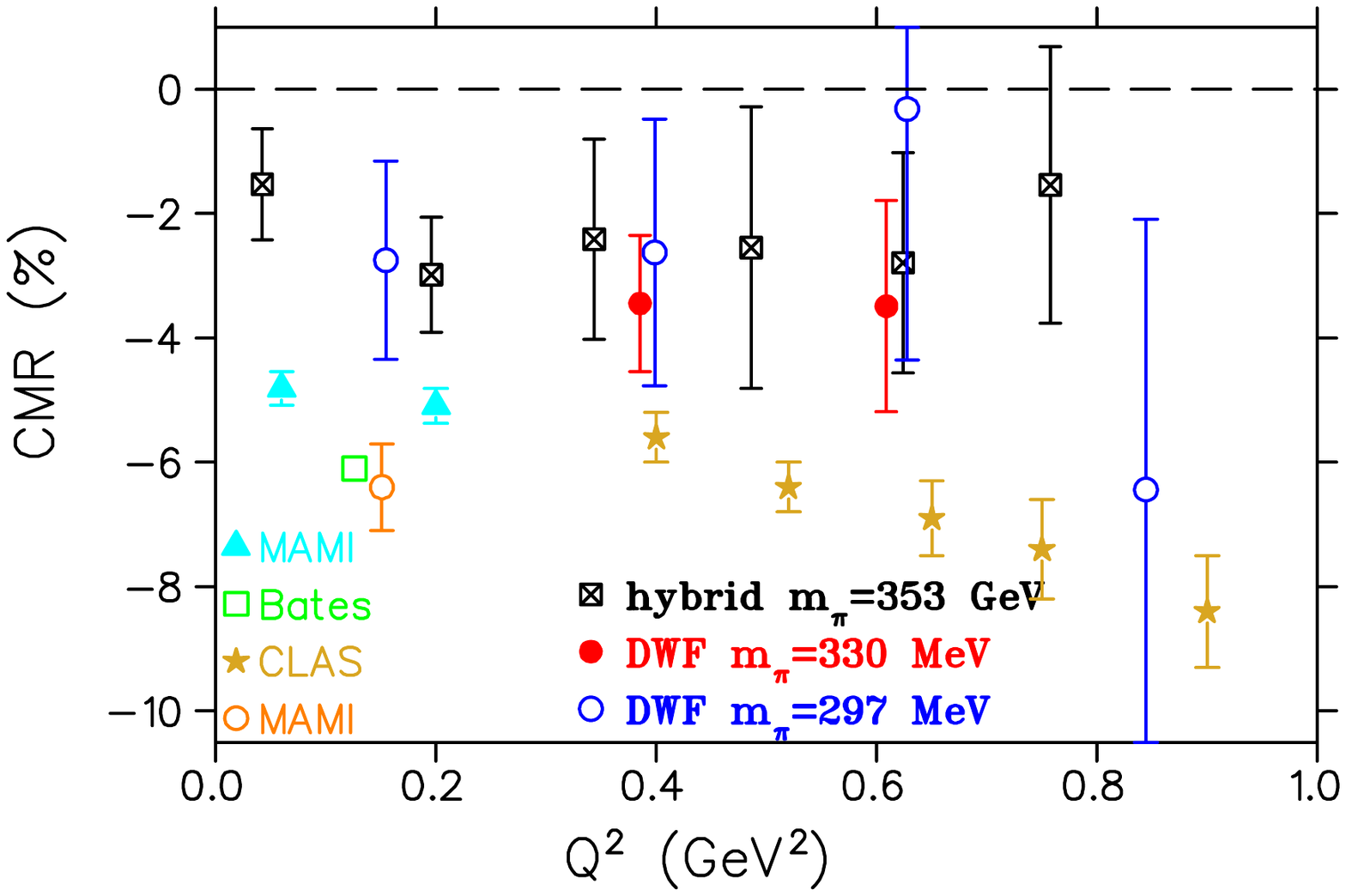}}
\end{minipage}}\vspace*{-4cm}
\caption{$N$ to $\Delta$ EM transition: Left: The transverse-longitudinal response function $\sigma_{LT}$ vs c.m. angle between p and $\gamma^*$ (from MAMI and Bates)~\cite{CNP}; The $N$ to $\Delta$ magnetic dipole FF (middle) and the ratio of Coulomb quadrupole to magnetic dipole FF (right) for the hybrid action and $N_f=2+1$ DWF.}
\label{fig:ND EM}
\end{figure}

$\bullet${ \bf $N\gamma^* \rightarrow \Delta$ form factors:}
There is an extensive experimental program to study
the $N$ to $\Delta$ EM transition and in particular to 
obtain accurate results on the sub-dominant quadrupole FFs
 ${ G}^*_{E2}(q^2)$ and ${G}^*_{C2}(q^2)$ that probe deformation. 
The experimental results, shown in Fig.~\ref{fig:ND EM}, are compatible with
 the
blue band obtained assuming deformation in the $N$-$\Delta$ and incompatible with the red band that includes no deformation.
These FFs can be computed within lattice QCD and since
 no disconnected contributions are involved  they provide
yet another benchmark for lattice methods.  
 The matrix element for $N$ to $\Delta$ EM  transition is written in terms of
three Sachs FFs as:
$$
{\scriptsize \langle \; \Delta (p',s') \; | j^\mu | \; N (p,s) \rangle=
{i\cal A}  \bar{u}_{\Delta,\sigma} (p',s')  \Biggl[ {G}^*_{M1}(Q^2) K^{\sigma \mu}_{M1} 
+{ {G}^*_{E2}}(Q^2) K^{\sigma \mu}_{E2} 
+{{G}^*_{C2}}(Q^2) K^{\sigma \mu}_{C2} \Biggr] u_N(p,s) \; \nonumber
}\,,$$
where ${\cal A}=\sqrt{\frac{2}{3}}\left({m_N m_\Delta}/{E_\Delta(\vec {p}^\prime) E_N(\vec{p})}\right )^{1/2}$ is a kinematical factor.

The
extraction of the sub-dominant quadrupole FFs is enabled
  by constructing optimized sources that isolate $ { G}^*_{E2}$ and   $ {G}^*_{C2}$~\cite{ND-Alexandrou}.
In Fig.~\ref{fig:ND EM} we show results 
using a hybrid action of DWF on staggered sea  as well as
 $N_f=2+1$ DWF, provided by RBC-UKQCD. For the dominant dipole FF $G^*_{M1}$, like for the nucleon FFs, we observe a
weaker $Q^2$-dependence as compared to experiment, that again highlights
the need of studying these FFs using simulations with $m_\pi< 300$~MeV. Lattice results,
shown in Fig.~\ref{fig:ND EM} for the ratio of the Coulomb quadrupole to
the magnetic dipole FF are 
non-zero. This is also true for the electric quadrupole FF,
pointing to a deformation of the $N$-$\Delta$.

\begin{figure}[h]
\begin{minipage}{0.33\linewidth}
\hspace*{-0.7cm}{\includegraphics[width=1.25\linewidth]{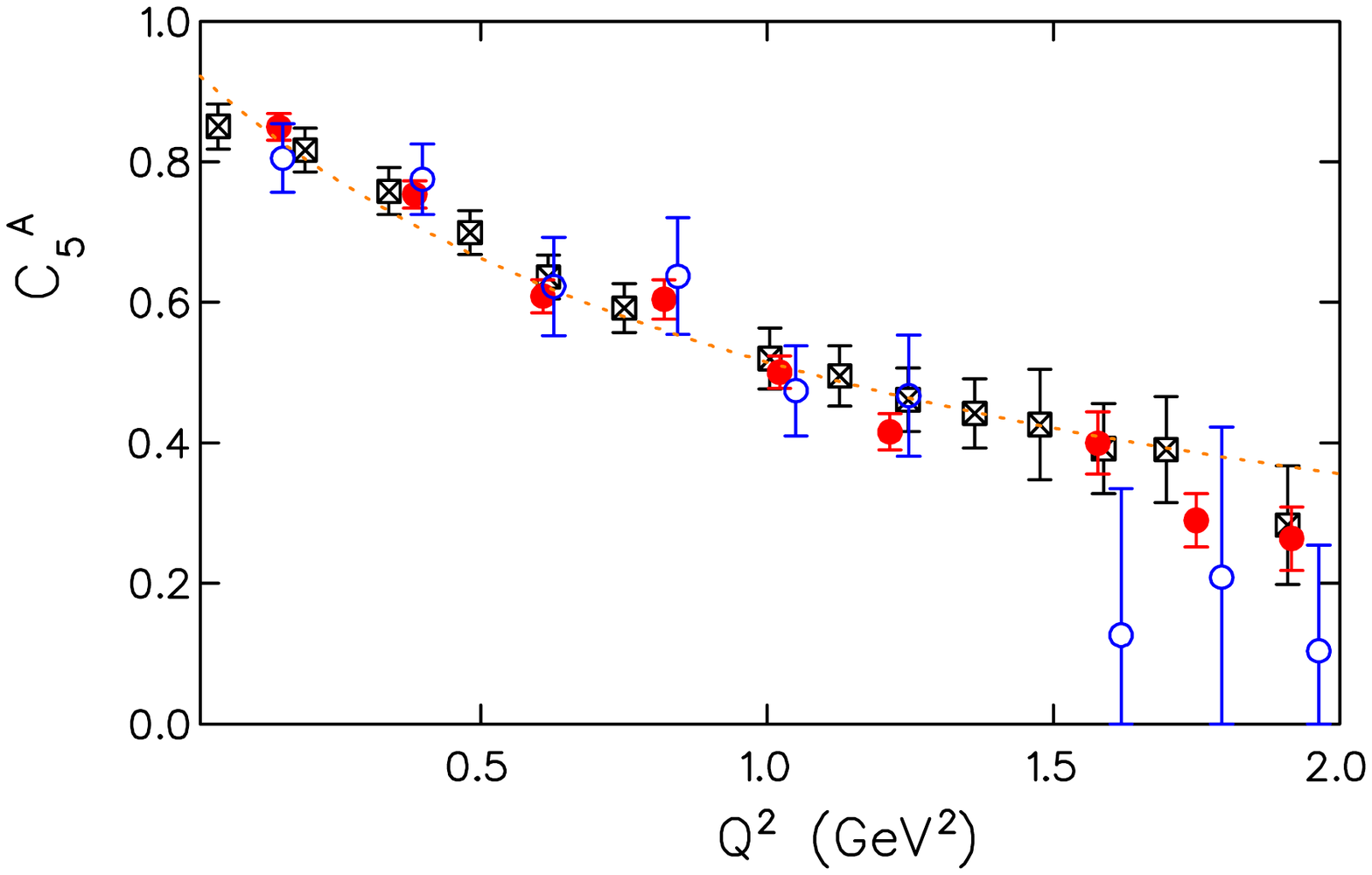}}
\end{minipage}\hfill
\begin{minipage}{0.33\linewidth}
\hspace*{-0.5cm}{\includegraphics[width=1.25\linewidth]{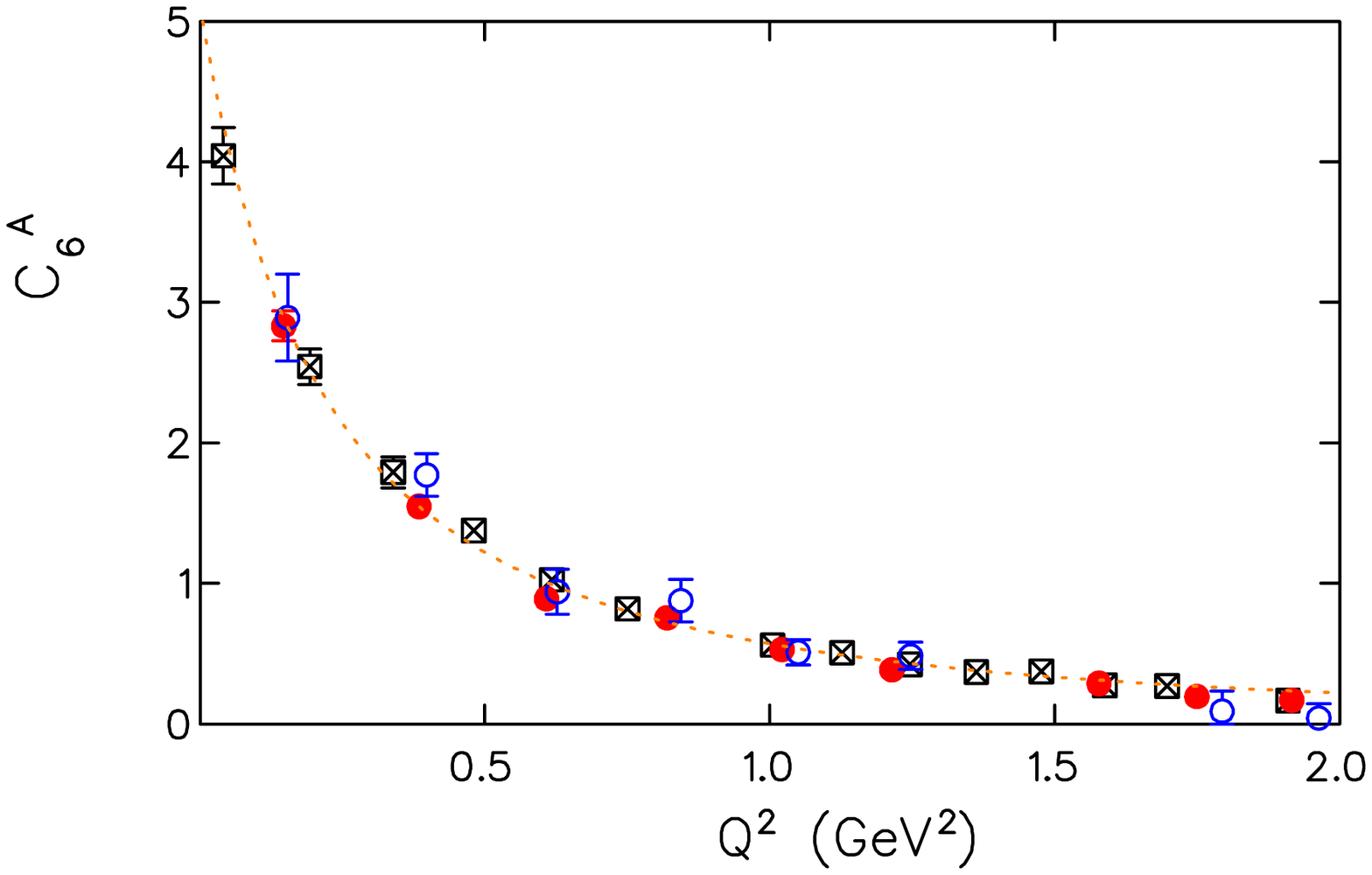}}
\end{minipage}\hfill
\begin{minipage}{0.33\linewidth}\vspace*{-3.5cm}
\hspace*{-0.3cm}{\includegraphics[width=1.25\linewidth, height=1.25\linewidth]{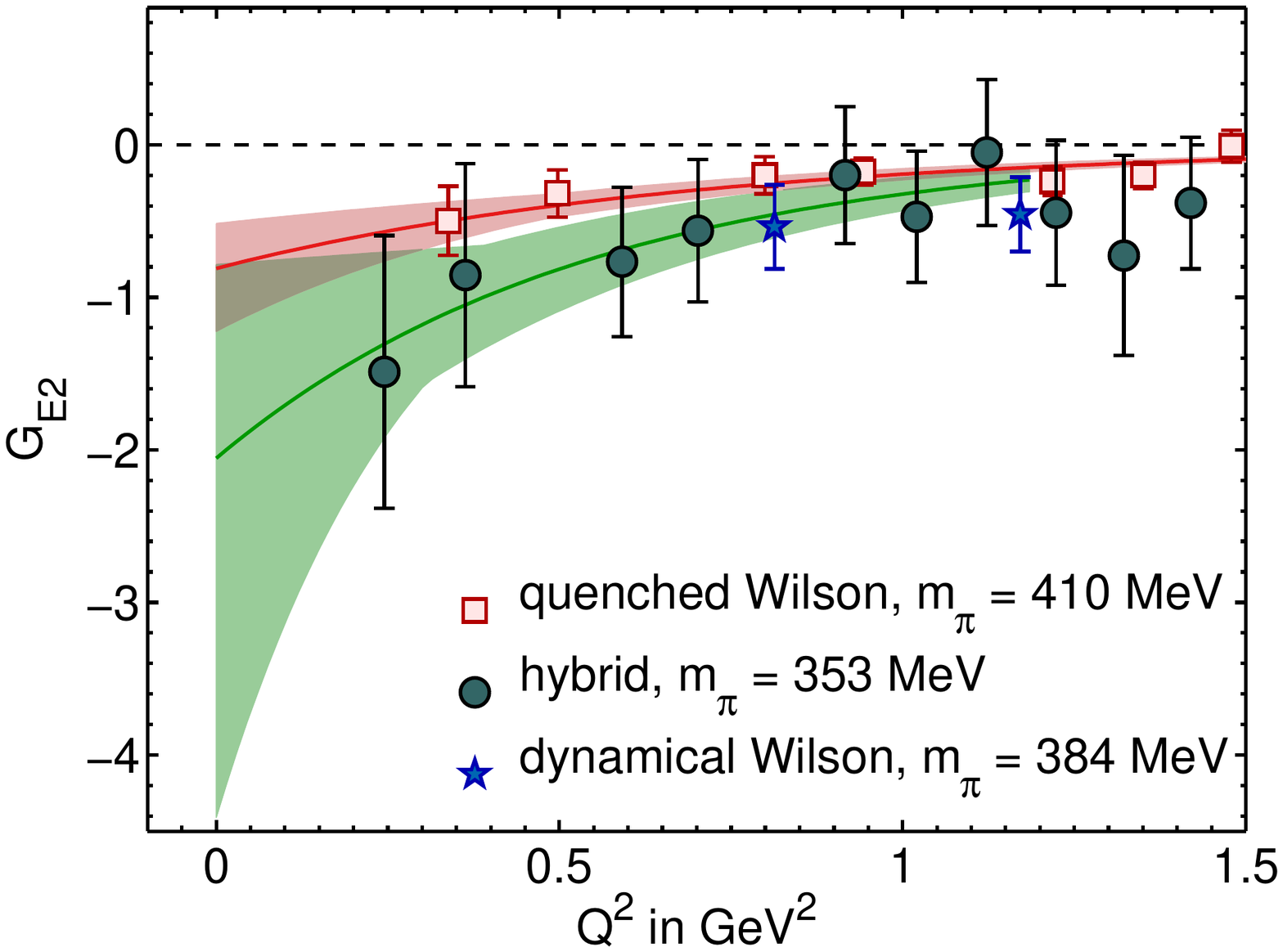}}
\end{minipage}
\vspace*{-4.0cm}
\caption{Left and middle: Axial $N$ to $\Delta$ FFs $C^A_5$ and $C^A_6$ respectively. The squares are for the hybrid action  at $m_\pi\sim 350$~MeV, the filled red circles for DWF at $m_\pi\sim 330$~MeV and the open blue circles for DWF at $m_\pi=300$~MeV. Right:
$\Delta$ electric quadrupole  FF for quenched, $N_f=2$ Wilson and $N_f=2+1$ hybrid 
action.}
\label{fig:NDaxial}
\end{figure}

$\bullet$ { \bf Weak N to $\Delta$ transition:}
In contrast to the EM transition, the weak $N$ to $\Delta$ is not well
studied experimentally.
 Therefore a lattice determination of
the axial vector $N$ to $\Delta$ FFs  would provide important input for phenomenology 
and in particular for 
chiral perturbation expansions. 
The weak N to $\Delta$ matrix element $\langle\Delta(p',s')|A_\mu^3|N(p,s)\rangle = \bar{u}_\Delta^\lambda(p',s') {\cal O}_{\lambda\mu} u_N(p,s)$ with
$$\small
{\cal O}_{\lambda\mu} = i{\cal A}
\Bigg[\left(\frac{ {C^A_3(Q^2)}}{m_N}\gamma^\nu +
            \frac{{C^A_4(Q^2)}}{m_N^2}p'^\nu\right)
            (g_{\lambda\nu}g_{\rho\nu}-g_{\lambda\rho}g_{\mu\nu})q^\rho 
           +{ C^A_5(Q^2)}g_{\lambda\mu}+\frac{{ C^A_6(Q^2)}}{m_N^2}q_\lambda
            q_\mu\Bigg]\,, \nonumber
$$
 where ${ C^A_5(Q^2)}$ is the equivalent of the nucleon FF $G_A(Q^2)$ and ${C^A_6}(Q^2)$ of  $G_p(Q^2)$  showing a pion pole behavior~\cite{ND-Alexandrou}. 
In Fig.~\ref{fig:NDaxial} we show results on the dominant FFs  $C^A_5$
and  $C^A_6$ obtained using the hybrid action   at $m_\pi\sim 350$~MeV and with $N_f=2+1$ 
DWF at $m_\pi\sim 330$~MeV and $m_\pi=300$~MeV. 
\vspace*{-0.8cm}

\vspace*{-0.5cm}

\section{$\Delta$ electromagnetic form factors and structure}\vspace*{-0.3cm}
Experimentally the $\Delta$ FFs are very difficult to measure due to the fact that the $\Delta$ decays strongly. Only its magnetic moment is measured experimentally albeit with a
large error.Therefore lattice calculations can complement experiment by
providing these FFs. 
\begin{figure}[h]
\begin{minipage}{0.33\linewidth} \vspace*{0cm}
\hspace*{-0.5cm}{\includegraphics[width=\linewidth]{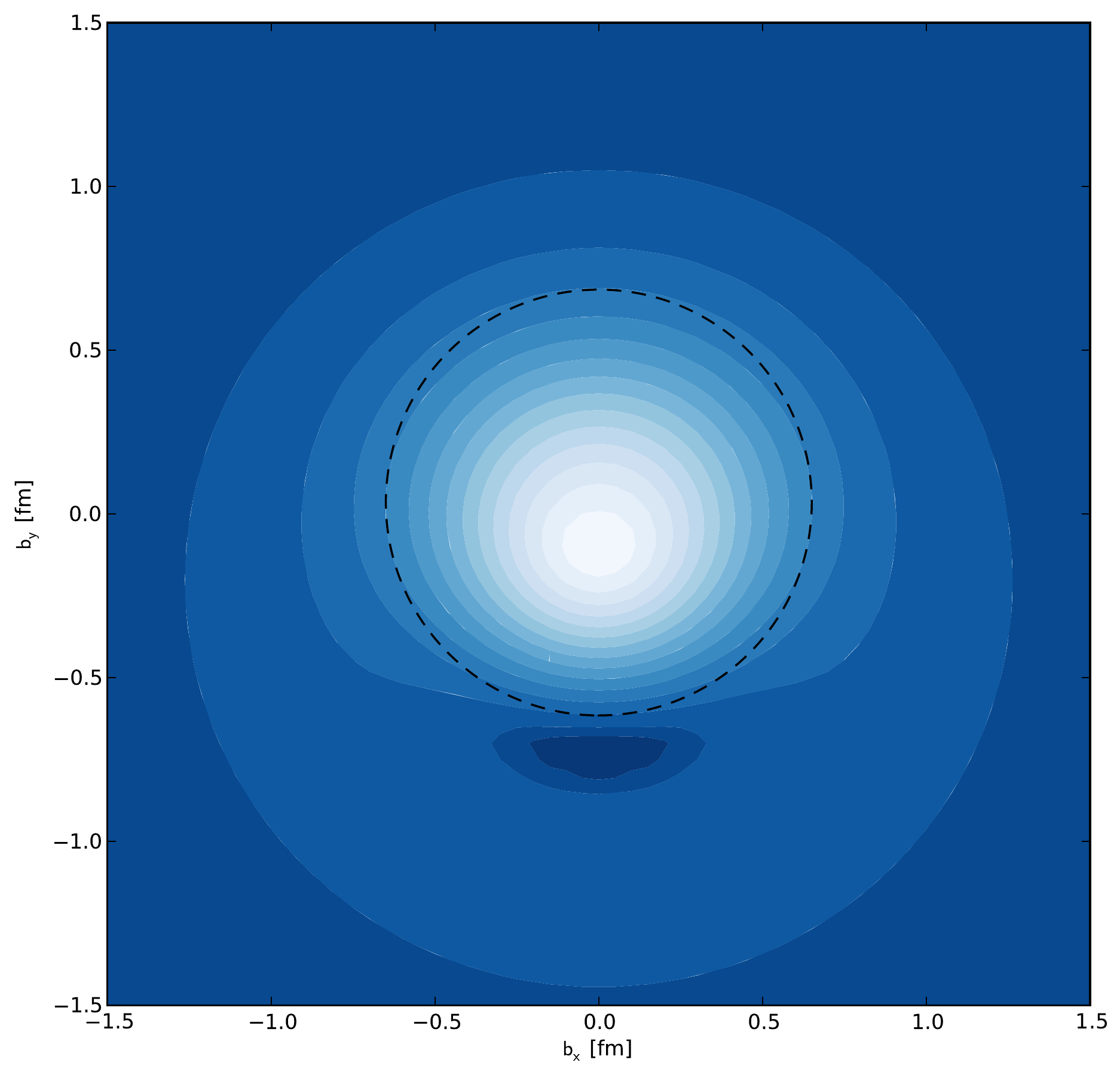}}
\end{minipage}\hfill
\begin{minipage}{0.33\linewidth}\vspace*{0cm}
\hspace*{-0.5cm}{\includegraphics[width=\linewidth]{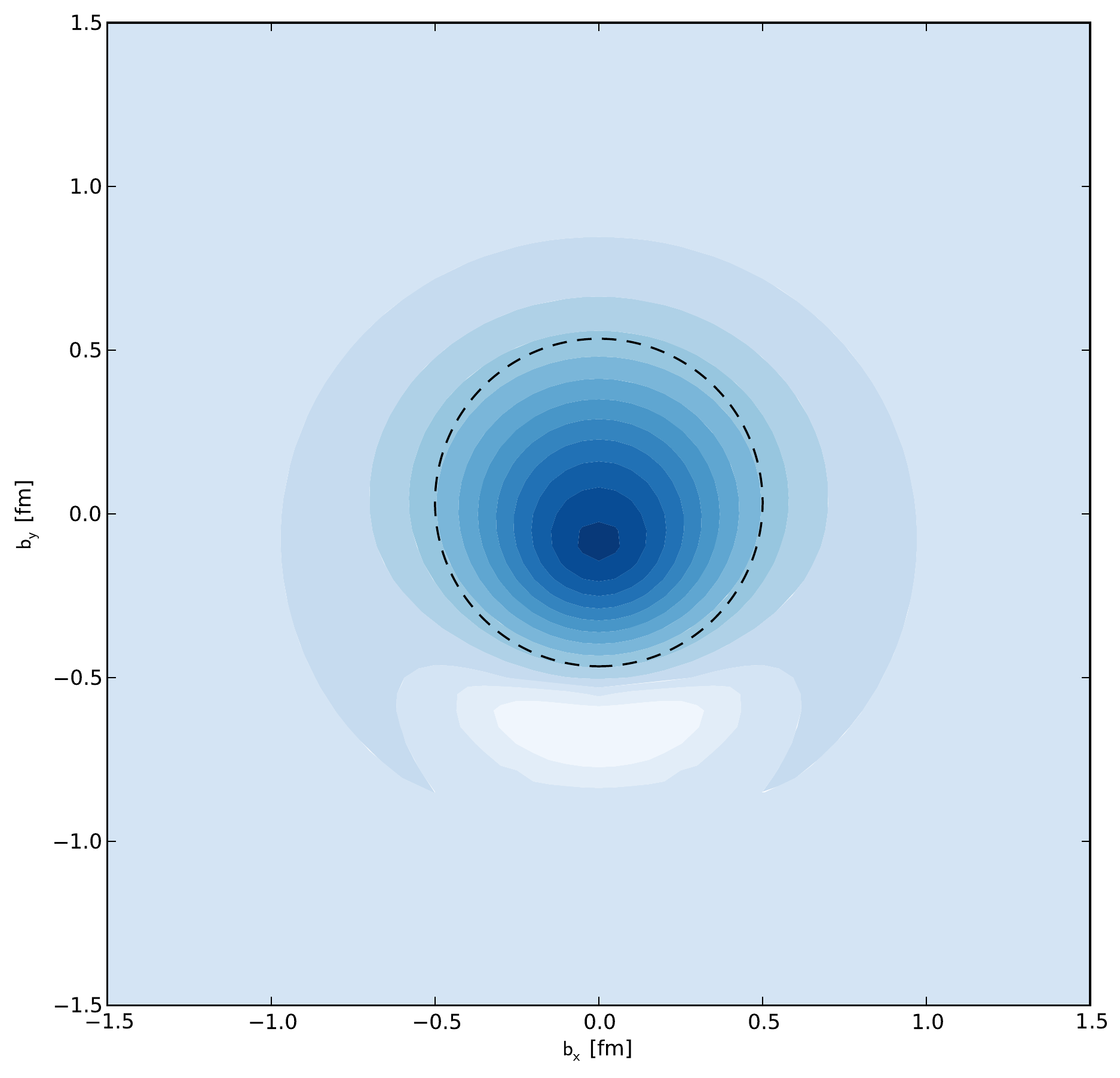}}
\end{minipage}\hfill
\label{fig:DD}
\begin{minipage}{0.33\linewidth}\vspace*{0cm}
\small{{\bf Figure 16:} Quark transverse charge densities in the $\Delta^{++}$ (left) and
$\Omega^-$ (right) for the 3/2-spin projection along the x-axis. Darker colors denote smaller values and  the charge of the particle is taken into account.
A dotted circle  of radius 0.5~fm is included for comparison. }
\end{minipage}
\end{figure}

\noindent
 The matrix element $\langle \Delta(p',s^\prime) |j^\mu(0) |\Delta (p,s)\rangle=\bar u_{\Delta,\alpha} (p',s^\prime) {\cal O}^{\alpha\mu\beta} u_{\Delta,\beta}(p,s)$ can be written  as

$$
{\cal O}^{\alpha\mu\beta}= -  {\cal A}_\Delta\left\{  \left[
F_1^\ast(Q^2)  g^{\alpha \beta}
+ F_3^\ast(Q^2) \frac{q^\alpha q^\beta}{(2 m_\Delta)^2}
\right] \gamma^\mu + \left[ F_2^\ast(Q^2)  g^{\alpha \beta}
+ F_4^\ast(Q^2) \frac{q^\alpha q^\beta}{(2m_\Delta)^2}\right]
\frac{i \sigma^{\mu\nu} q_\nu}{2 m_\Delta} \, \right\} \nonumber
$$
with e.g. the quadrupole FF given by: 
 $ G_{E2} = \left( F_1^\ast - \tau F_2^\ast \right) - \frac{1}{2} ( 1 + \tau)
\left( F_3^\ast - \tau F_4^\ast \right)$, where $\tau \equiv Q^2 / (4 m_\Delta^2)$.
Optimized sources are constructed to isolate the quadrupole FF $G_{E2}$,
which probes deformation. The transverse charge density of a $\Delta$ polarized along the x-axis can be defined in the infinite momentum frame.
Using $G_{E2}$ we can predict the `shape' of $\Delta$. The result is shown in
Fig.~16 and for spin projection $3/2$ it is elongated along the spin axis.
The $\Omega^-$ shows a similar but smaller deformation~\cite{DD-Alexandrou}.
The weak $\Delta$ FFs can be computed in an analogous manner~\cite{eric}.

\section{Conclusions}
The nucleon EM form factors provide a benchmark for lattice QCD beyond hadron masses. 
 Most collaborations obtain results for the isovector FFs up to about  $Q^2=2$~GeV$^2$. 
Systematic studies of  lattice artifacts  on GFFs are  now under way and
recent data reveal 
that cut-off effects are negligible for $a\stackrel{<}{\sim} 0.1$~fm, whereas finite volume corrections, although difficult to evaluate, are
within the current statistical errors of $\sim (2-3)$\% for $Lm_\pi \stackrel{>}{\sim} 3.3$. A possible
exception
is $ {G_p}$ at low $Q^2$-values.
We find that, in general, lattice results 
using different discretization schemes are
consistent
 but they show a milder $Q^2$-dependence as compared to experiment.
As illustrated in the case of the nucleon axial charge, the 
biggest uncertainty in comparing with experiment
 is the chiral extrapolation. Therefore
 a lattice determination of a number of couplings used as input in chiral extrapolations  will enable global fits to e.g. the $N$-$\Delta$ system that can
help extrapolation to the physical point.
 Interesting  questions such as the `shape' of a hadron can be addressed using input from lattice form factors
as demonstrated for the $\Delta$ and $\Omega$. 
Moments of  GPDs yield more detailed information on both longitudinal and transverse
distributions and a tomography of hadrons can be obtained by studying these quantities.
We therefore, conclude that, 
overall, there is  good progress in baryon structure calculations and that
we now are in an exciting era, having simulations
 close enough to the physical point,  
in order to probe interesting dynamics in hadronic systems.

\vspace*{0.32cm}
\noindent
 {\bf Acknowledgments:} 
I would like to thank the members of the ETM collaboration and in particular
M. Brinet, J. Carbonell, M. Constantinou, V. Drach, P. A. Harraud, K. Jansen, T. Korzec,
M. Papinutto and O. Pene, my long-term collaborators G. Koutsou, J. W. Negele,
and A. Tsapalis as well as Y. Proestos and  M. Vanderhaeghen 
 for their valuable input 
on the topics presented.
This work was performed using HPC resources from GENCI (IDRIS and CINES) Grant 2009-052271, the Blue-Gene/P at JSC
and was partly supported through funding received   by  the
 Cyprus Research Promotion Foundation under contracts EPYAN/0506/08,
KY-$\Gamma$/0907/11,  TECHNOLOGY/$\Theta$E$\Pi$I$\Sigma$/0308(BE)/17
and $\Delta$IE$\Theta$NH$\Sigma$/$\Sigma$TOXO$\Sigma$/0308/07.



\begin{thebibliography}{99}
\bibitem{Durr} St. D\"urr {\it et al.} (BMW), Science 322, 1224 (2008). 
\bibitem{Alexandrou} C. Alexandrou {\it et al.} (ETMC), Phys. Rev. D {\bf 80}, 114503 (2009). 
\bibitem{fpi}
T.~Kaneko {\it et al.} (JLQCD),
  PoS {\bf LAT2007}, 148 (2007); A.~J\"uttner {\it et al.} (RBC-UKQCD),
  PoS C {\bf D09} (2009) 010;   D.~Br\"ommel {\it et al.}  (QCDSF-UKQCD)
  Eur.\ Phys.\ J.\  C {\bf 51}, 335 (2007);  D. Br\"ommel {\it et al.} (QCDSF), Eur. Phys. J. C 51, 335 (2007); F. D. R. Bonnet {\it et al.},  Phys. Rev. D. {\bf 72}, 054506 (2005).
\bibitem{ETMC} R. Frezzotti, V. Lubicz and S. Simula, Phys. Rev. D {\bf 79}, 074506 (2009). 
\bibitem{rho-width} J. Frison {\it et al.}, arXiv:1011.3413; S. Aoki {\it et al.}  (PACS-CS),	arXiv:1011.1063; 
S.~Aoki {\it et al.},
  Phys.\ Rev.\  D {\bf 76}, 094506 (2007); M.~G\"ockeler {\it et al.} (QCDSF),
  PoS {\bf LATTICE2008}, 136 (2008).
\bibitem{Feng}
  X.~Feng, K.~Jansen and D.~B.~Renner,
  arXiv:0910.4871; arXiv:1011.5288.
\bibitem{LHPC} A. Waker-Loud {\it et al.} (LHPC), Phys. Rev. D {\bf 79} 054502 (2009). 
\bibitem{PACS} S. Aoki {\it et al.} (PACS-CS), Phys. Rev. D {\bf 79}, 034503 (2009).  
\bibitem{excited}
  J.~M.~Bulava {\it et al.},
  Phys.\ Rev.\  D {\bf 79}, 034505 (2009);
  M.~S.~Mahbub {\it et al.},
  Phys.\ Lett.\  B {\bf 679}, 418 (2009);
G.~Engel, {\it et al.},
  arXiv:0910.2802 [hep-lat].
\bibitem{Diehl} M. Diehl, Phys. Rep. 388, 41 (2003). \\
\bibitem{Ji} X. Ji, J.~Phys.~G24, 1181 (1998).
\bibitem{Zanotti}  J.~M.~Zanotti,
  PoS {\bf LATTICE2008}, 007 (2008).
\bibitem{Blossier}
  B.~Blossier, M.~Della Morte, G.~von Hippel, T.~Mendes and R.~Sommer,
  JHEP {\bf 0904}, 094 (2009).
\bibitem{Lin}  H.-W. Lin {\it et al.}, arXiv:1005:0799
\bibitem{Gockeler}   M. G\"ockeler {\it et al.} (QCDSF), Nucl. Phys. {\bf B544}, 699 (1999).
\bibitem{Z-Alexandrou}
  C.~Alexandrou, M.~Constantinou, T.~Korzec, H.~Panagopoulos and F.~Stylianou,
  arXiv:1006.1920.
\bibitem{Z2-Alexandrou}
  C.~Alexandrou, M.~Constantinou, T.~Korzec, H.~Panagopoulos and F.~Stylianou, PoS {\bf Lattice 2010}, 224 (2010); in preparation.
\bibitem{Aoki} Y. Aoki {\it al.} (RBC-UKQCD), Phys. Rev. D {\bf 82}, 014501 (2010).
\bibitem{FF-Bratt}   J. D. Bratt {\it et al.} (LHPC), arXiv:1001.3620.
\bibitem{FF-Alexandrou}     C. Alexandrou {\it et al.} (ETMC), PoS {\bf LAT2009}, 145 (2009), arXiv:0910.3309.
\bibitem{GPD-Zanotti} J. Zanotti (QCDSF), private communication.
\bibitem{Alexandrou-new} C. Alexandrou {\it et al.} (ETMC), in preparation.
\bibitem{FF-Yamazaki} T. Yamazaki {\it et al.} (RBC-UKQCD), Phys. Rev. D {\bf 79}, 114505 (2009). 
\bibitem{Khan} A. Ali Khan {\it et al.} (QCDSF), Phys. Rev. D {\bf 74}, 094508 (2006).
\bibitem{SSE} T. R. Hemmert, M. Procura and W. Weise, Phys. Rev. D {\bf 68}, 075009 (2003).
\bibitem{FF-Syritsyn} S. N. Syritsyn {\it et al.} (LHPC), Phys. Rev. D {\bf 81}, 034507 (2010).
\bibitem{Wittig}  S.~Capitani, M.~Della Morte, B.~Knippschild and H.~Wittig,
  arXiv:1011.1358 [hep-lat].
\bibitem{FF-Hemmert} T. R. Hemmert and W. Weise, Eur. Phys. J. A {\bf 15}, 487 (2002); M. G\"ockeler {\it et al.}, Phys. Rev. D {\bf 71}, 034508 (2005).
\bibitem{Hagler} M.~G\"ockeler {\it et al.}  (QCDSF-UKQCD),
  PoS {\bf LATTICE2008}, 138 (2008).
\bibitem{korzec} C. Alexandrou {\it et al.} (ETMC), PoS {\bf LAT2009}, 136 (2009); T. Korzec, private communication.
\bibitem{Dru} D.~B.~Renner, PoS {\bf LAT2009} (2009),
  arXiv:1002.0925.
\bibitem{Thomas}  D. Arndt, M. Savage, Nucl. Phys. {\bf A697}, 429 (2002);  W.~Detmold, W~Melnitchouk, A.~Thomas, Phys. Rev. D {\bf 66}, 054501 (2002).
\bibitem{CNP} C. N. Papanicolas, Eur. Phys. J. A{\bf 18}, 141 (2003).
\bibitem{ND-Alexandrou}
C. Alexandrou {\it et al.}, PoS {\bf LAT2009}, 156 (2009), arXiv:0910.5617; C.Alexandrou, Th. Leontiou, J. W. Negele and A. Tsapalis, Phys. Rev. Lett. 98, 052003 (2007).
\bibitem{DD-Alexandrou}  C. Alexandrou {\it et al.}, Phys. Rev. D {\bf 79}, 014507 (2009); Nucl. Phys. {\bf A825}, 115 (2009).
\bibitem{eric} C. Alexandrou {\it et al.}, PoS {\bf Lattice 2010}, 141 (2010), arXiv:1011.3233.
\end{thebibliography}
\end{document}